\documentclass[aps,preprint,epsfig,rotate]{revtex4}
\usepackage{graphicx}
\usepackage{bm}
\usepackage{epsfig}

\input epsf

\begin{document}
%
\title{On Maxwell electrodynamics in multi-dimensional spaces}

 \author{Alexei M. Frolov}
 \email[E--mail address: ]{alex1975frol@gmailcom}

\affiliation{Department of Applied Mathematics, \\
 University of Western Ontario, London, Ontario N6H 5B7, Canada}

\date{\today}

\begin{abstract}

The governing equations of Maxwell electrodynamics in multi-dimensional spaces are derived from the 
variational principle of least action which is applied to the action function of the electromagnetic 
field. The Hamiltonian approach for the electromagnetic field in multi-dimensional pseudo-Euclidean 
(flat) spaces has also been developed and investigated. Based on the two arising first-class 
constraints we have generalized to multi-dimensional spaces a number of different gauges known for 
the three-dimensional electromagnetic field. For multi-dimensional spaces of non-zero curvature the 
governing equations for the multi-dimensional electromagnetic field are written in manifestly 
covariant form. Multi-dimensional Einstein's equations of metric gravity in the presence of 
electromagnetic field have been re-written in the true tensor form. Methods of scalar electrodynamics
are applied to analyze Maxwell equations in the two- and one-dimensional spaces. 

\noindent
This version is close to the final version published in: Universe {\bf 8}, 20 (2022). 

\end{abstract}

\maketitle


\section{Introduction}

The main goal of this communication is to develop the logically closed and non-contradictory version of 
electrodynamics in the multi-dimensional (or $n-$dimensional) space. Right now, such a development can 
be considered as a pure theoretical (or model) task, but originally, our plan was to include the 
multi-dimensional electromagnetic fields in our Hamiltonian analysis of the metric gravity 
\cite{Fro2021}. Note that all Hamiltonian approaches which  are based on the $\Gamma - \Gamma$ Lagrangian 
(see, e.g., \cite{Fro2021} and earlier references therein) have been derived in the manifestly covariant
form and can be applied to multi-dimensional (or $n-$dimensional, where $n (\ge 3)$ is an arbitrary 
integer) Riemannian spaces without any modification. On the other hand, our current Maxwell theory of 
electromagnetic field and corresponding Hamiltonian approach can be used only for three-dimensional 
(geometrical) spaces. This contradiction creates numerous problems for the development of any united 
theory of the coupled electromagnetic and gravitational fields. Also, it is hard to believe that in 
reality one can smoothly combine two theories that have different properties with respect to their 
extensions on multi-dimensional spaces. 

After our investigations have begun it did not take long to understand that such a theory of the free 
electromagnetic fields in multi-dimensions simply does not exist even in the first-order approximation 
(in contrast with the metric gravity). There are quite a few reasons why similar generalization of the 
classical electrodynamics to multi-dimensional spaces has not been developed earlier. For instance, the 
explicit expression for the action integral and, therefore, for the Lagrangian of the electromagnetic 
field in multi-dimensions is unknown. However, if we do not know the Lagrangian of multi-dimensional 
electromagnetic filed, then it is impossible to construct any valuable Hamiltonian. There were a number 
of smaller problems which substantially complicated any direct generalization of Maxwell theory to 
$n-$dimensional spaces. One of them is the lack of a reliable and practically valuable definition of a 
$curl-$operator (or $rot$-operator) in multi-dimensional spaces, where $n \ge 4$. In general, it is 
tricky to develop multi-dimensional electrodynamics without such an operator. Finally, we have decided 
to investigate this problem and derive some useful results which are of great interest for the 
Hamiltonian formulation of the metric gravity combined with electromagnetic field(s) in multi-dimensional 
spaces. 

First, let us briefly discuss the classical Maxwell equations known for the three-dimensional 
electromagnetic fields. For the first time, the Maxwell equations were written by J.C. Maxwell 
in 1862 (published in 1865 \cite{Maxw} (see also \cite{Maxw1} and \cite{Fro2015})) for the 
intensities of electric ${\bf E}$ and magnetic ${\bf H}$ fields (or for the electric and magnetic 
field strengths):
\begin{eqnarray}
 div {\bf E} &=& 4 \pi \rho \; \; , \; \; curl {\bf E} = -\frac{1}{c} \frac{\partial 
 {\bf H}}{\partial t}  \; \; \; , \nonumber \\
 div {\bf H} &=& 0 \; \; , \; \; curl {\bf H} = \frac{1}{c} \frac{\partial {\bf E}}{\partial t} + 
 \frac{4 \pi}{c} {\bf j} \; \; , \; \label{Maxw3D} 
\end{eqnarray}
where $\rho$ and ${\bf j} = \rho {\bf v}$ are the electric charge density (scalar) and electric current 
density (vector), respectively. In this study the charge density and current are defined exactly as in 
\$ 29 from \cite{LLTF}. Later, it was noticed by Hertz and others that these four equations from 
Eq.(\ref{Maxw3D}) can be re-written in a simple form, if we can introduce the four-dimensional potential 
$\bar{A} = (\varphi, {\bf A})$, where $\varphi$ is the scalar potential and ${\bf A}$ is the vector 
potential of the electromagnetic field. Note that the scalar potential $\varphi$ can equally be considered 
as 0-component ($A_0$) of the four-dimensional vector potential $\bar{A}$ of the electromagnetic field. 
The $\varphi$ and ${\bf A}$ potentials are simply related to the intensities of electric ${\bf E}$ and 
magnetic ${\bf H}$ fields: ${\bf H} = curl {\bf A}$ and ${\bf E} = - \frac{\partial {\bf A}}{\partial t} 
- grad \varphi$. By using these relations between the potentials ($\varphi, {\bf A}$) and intensities 
$({\bf E}, {\bf H})$ of electromagnetic field one finds that the second equation in the first line and 
first equation in the second line of Eq.(\ref{Maxw3D}) hold identically. The two remaining equations from 
Eq.(\ref{Maxw3D}) lead to the following non-homogeneous equations
\begin{eqnarray}
 & &\frac{1}{c^{2}} \frac{\partial^{2} {\bf A}}{\partial t^{2}} - \Delta {\bf A} + grad \Bigl( div {\bf 
 A} + \frac{1}{c} \frac{\partial \varphi}{\partial t} \Bigr) = \frac{4 \pi}{c} {\bf j} \; \; \label{M4} \\
  &-&\Delta \varphi - \frac{1}{c} div \Bigl(\frac{\partial {\bf A}}{\partial t}\Bigr) = 4 \pi \rho \; \; 
  \label{3M} 
\end{eqnarray}
where $\Delta = \frac{\partial^2}{\partial x^{2}} + \frac{\partial^2}{\partial y^{2}} + 
\frac{\partial^2}{\partial z^{2}}$ is the three-dimensional Laplace operator. By applying the 'gauge 
condition' $\frac{\partial \varphi}{\partial t} + div {\bf A} = 0$ for the four-dimensional potential, 
one reduces the two last equations to the form 
\begin{eqnarray}
 && \frac{1}{c^{2}} \frac{\partial^{2} {\bf A}}{\partial t^{2}} - \Delta {\bf A} = \frac{4 \pi}{c} 
 \rho {\bf v} \; \; , \; \label{M4a} \\
 && \frac{1}{c^{2}} \frac{\partial^{2} \varphi}{\partial t^{2}} - \Delta \varphi  = 4 \pi \rho \; \; , 
 \; \; \label{3Ma} 
\end{eqnarray}
where the operator $\frac{1}{c^{2}} \frac{\partial^{2}}{\partial t^{2}} - \Delta$ is the four-dimensional 
Laplace operator in pseudo-Euclidean space, which is often called the $d^{,}$Alembertian operator.  

It is interesting that all equations mentioned above can be derived by varying the action functional $S$ 
which is written for a system of particles and electromagnetic field(s) interacting with these particles. 
In Gauss units the explicit form of this action function (or action, for short) $S$ is 
\begin{eqnarray}
 S = S_p + S_{fp} + S_f = - \sum_{k} \int m_k c ds_k - \sum_{k} \int \frac{e_k}{c} A_{\alpha}(k) 
 dx^{\alpha} - \frac{1}{16 \pi} \int F_{\alpha\beta} F^{\alpha\beta} d\Omega \; \; , \; \label{actionX} 
\end{eqnarray}
where the two sums are taken over particles, $s = \sqrt{x_{\mu} x^{\mu}} = \sqrt{g_{\mu\nu} x^{\mu} 
x^{\mu}}$ is the interval, $S_p$ is the action for the particles ($k = 1, 2, \ldots$), $S_{fp}$ is the 
action which describes interaction between particles and electromagnetic field, while $S_f$ is the action 
for the electromagnetic field itself. The notation $e_k$ stands for the electric charge of the $k$-th 
particle, while $m_k$ means the mass of the same particle and $A_{\alpha}$ is the covariant component of 
the four-dimensional vector potential $\bar{A}$ of the electromagnetic field. This formula, 
Eq.(\ref{actionX}), is written for the four-dimensional pseudo-Euclidean (flat) space-time. This fact 
drastically simplify analysis and derivation of the Maxwell and other equations in classical 
three-dimensional electrodynamics.   

In this study we discuss a possibility to generalize the usual (or three-dimensional) Maxwell equations 
to the spaces of larger dimensions. In respect to this, below we shall consider $n-$dimensional, pure 
geometrical spaces and $(n + 1)-$dimensional space-time manifolds. Our main goal is to derive the correct 
form of multi-dimensional Maxwell equations and investigate their basic properties. In particular, we 
want to understand how many and what kind of changes can we expect in the multi-dimensional Hamiltonian 
of the free electromagnetic field and in a number of arising first-class constraints. A separate, but 
closely related problem is the gauge invariance of the free electromagnetic field. Another interesting 
problem is to investigate the explicit form of multi-dimensional Maxwell equations in the presence of 
multi-dimensional gravitational fields. A brief discussion of scalar electrodynamics can be found in 
Appendix A. All new results obtained in the course of our current analysis will be used later to develop 
the modern united theory of electromagnetic and gravitational fields.   

\section{Scalar and vector potentials of the electromagnetic field} 

Let us derive the closed system of Maxwell equations for the $n-$dimensional (geometrical) space, where $n \ge 3$. 
The time $t$ is always considered as an independent scalar and special $(n + 1)$-st variable. This means that we 
are dealing with manifolds of variables defined in $(n + 1)-$dimensional space-time. First, we need to define the 
vector potential $\bar{A}$ in this $(n + 1)-$dimensional space-time. Based on experimental facts known for actual 
electromagnetic systems considered in one, two and three-dimensions, below we shall assume that interaction of a 
point particle with the electromagnetic field is determined by a single, scalar parameter $e$, which is the electric 
charge of this particle. The parameter $e$ can be positive, negative, or equal zero. The properties of the 
electromagnetic field are described by the $(n + 1)$-dimensional vector potential $\bar{A}$. The notation $A_{\mu}$ 
(or $\bar{A}_{\mu}$) stands for the covariant $\mu-$component of this $(n + 1)$-dimensional vector potential 
$\bar{A}$. In this study we also deal with $n-$dimensional space-like vector potential ${\bf A}$. Co- and 
contravariant components of this vector are designated by Latin indexes, e.g., $A_{k}$ and $A^{k}$, where $k = 1, 
2, \ldots, n$. The same rule is applied to all vectors and tensors mentioned in this study: components of $(n + 
1)-$vectors are labelled by Greek indices (each of which varies between 0 and $n$), while spatial components of 
these $n-$dimensional vectors (each varies between 1 and $n$) are denoted by Latin indices. Generalization of this 
rule to the tensors of arbitrary ranks is straightforward and simple. Note also that in all formulas below the 
following 'summation rule' is applied: a repeated suffix (or index) in any formula means summations over all values 
of this suffix (or index).  

In general, the vector potential $\bar{A}$ can be written in the form $\bar{A} = (\varphi, {\bf A})$, which 
includes the scalar potential $\varphi (= A_0)$ and $n-$dimensional vector potential ${\bf A} = (A_1, A_2, \ldots, 
A_n)$. For arbitrary scalar $\Phi$ and vector ${\bf V}$ functions in $n-$dimensional space we can determine the 
first-order differential operators: (a) gradient operator $\nabla$ (or $grad$) and (b) divergence operator $div$. 
They are defined as follows 
\begin{eqnarray}
 \nabla \Phi = grad \; \Phi = \Bigl( \frac{\partial \Phi}{\partial x_1}, \frac{\partial \Phi}{\partial x_2}, \ldots, 
 \frac{\partial \Phi}{\partial x_n} \Bigr) \; \;  {\rm and} \; \; div \; {\bf V} = \frac{\partial V_1}{\partial x_1} 
 + \frac{\partial V_2}{\partial x_2} + \ldots + \frac{\partial V_n}{\partial x_1} \; \; \; \label{divgrad}
\end{eqnarray}
Analogous definitions of these two operators can easily be generalized and applied to the scalar and vector 
functions defined in $(n + 1)-$dimensional space. By using these definitions we can discuss the gradient of 
the scalar potential $\nabla \varphi (= \nabla A_0)$ (vector) and divergence of vector potential $div {\bf A}$ 
(scalar) in the $n-$dimensional space.   

The $(n + 1)-$dimensional vector potential $\bar{A} = (A_0, A_1, \ldots, A_n)$ allows us to define the truly 
antisymmetric $(n + 1) \times (n + 1)$ electromagnetic field tensor $F_{\alpha\beta} (= - F_{\beta\alpha})$ by using 
the relation 
\begin{eqnarray}
 F_{\alpha\beta} = \frac{\partial A_{\beta}}{\partial x^{\alpha}} - \frac{\partial A_{\alpha}}{\partial 
 x^{\beta}} = - F_{\beta\alpha} \; \; , \; \; {\rm and} \; \; F^{\alpha\beta} = \frac{\partial A^{\beta}}{\partial 
  x_{\alpha}} - \frac{\partial A^{\alpha}}{\partial x_{\beta}} = - F^{\beta\alpha} \; , \; \label{tensor} 
\end{eqnarray}
which formally coincides with the analogous definition of this tensor known in the four-dimensional space-time. For 
$(n + 1)-$dimensional space-time manifold this tensor has zero-diagonal matrix elements (or components), i.e., 
$F_{\alpha\alpha} = 0$. Therefore, in $n-$dimensional space each of the antisymmetric $F^{\alpha\beta}$ and 
$F_{\alpha\beta}$ tensors have $\frac{n (n - 1)}{2}$ different and independent components. The double sum 
$F_{\alpha\beta} F^{\alpha\beta}$ is the first (or main) invariant of the electromagnetic field defined in the 
$(n + 1)$-dimensional space. Now, let us write the following explicit formula for the action $S$ for the system, 
which includes the particles and electromagnetic field itself. This action takes the following form (see, e.g., 
\cite{LLTF}) 
\begin{eqnarray}
 S = S_p + S_{fp} + S_f = - \sum_{k} \int m_{k} c ds - \sum_{k} \int \frac{e_{k}}{c} A_{\alpha}(k) dx^{\alpha} - 
 a \int F_{\alpha\beta} F^{\alpha\beta} d\Omega \; \; , \; \label{actionu} 
\end{eqnarray}
where $s = \sqrt{x_{\mu} x^{\mu}} = \sqrt{g_{\mu\nu} x^{\mu} x^{\mu}}$ is the interval, $S_p$ is the action 
function for the particles, $S_{fp}$ is the action function which describes interaction between particles and 
electromagnetic field and $S_f$ is the action function for the electromagnetic field itself. In this equation 
the summation is performed over all particles (index $k$). The notation $A_{\alpha}(k)$ that the 
$\alpha-$component of the vector potential must be determined at the point of location of $k-$th particle. 
Note that the formula, Eq.(\ref{actionu}), is applicable in the flat pseudo-Euclidean and/or Euclidean spaces 
only. Its generalization to multi-dimensional Riemannian spaces (spaces of non-zero curvature) is considered 
below. At the next step we need to determine the constant $a$ in Eq.(\ref{actionu}). This can be achieved by 
considering the Coulomb's law in multi-dimensions (see, the next Section).  

In conclusion of this Section we want to emphasize the fact that our action function, which is chosen in the 
form of Eq.(\ref{actionu}), allows one to derive the equations of motion for a system of electrically charged,
point particles which move in the electromagnetic field. For instance, for one electrically charged particle 
by varying the coordinates of this particle (i.e., the $x^{\mu}$ and $x^{\alpha}$ variables) in the action 
function, Eq.(\ref{actionu}), one finds the following equation of motion for one electrically charged, point 
particle which moves in the non-flat multi-dimensional space 
\begin{eqnarray}
 \frac{d^{2} x^{\alpha}}{ds^{2}} + \Gamma^{\alpha}_{\beta\gamma} \frac{d x^{\beta}}{ds} \frac{d x^{\gamma}}{ds}
 - \frac{e}{c} F^{\alpha\beta} g_{\beta\gamma} \frac{d x^{\gamma}}{ds} = 0 \; \; , \; {\rm or} 
 \; \; \frac{d^{2} x^{\alpha}}{ds^{2}} + \Gamma^{\alpha}_{\beta\gamma} \frac{d x^{\beta}}{ds} 
 \frac{d x^{\gamma}}{ds}- \frac{e}{m c^{2}} F^{\alpha}_{\beta} \frac{d x^{\beta}}{ds} = 0  \; \; , \; \; 
 \label{eqamot}
\end{eqnarray}
where $\Gamma^{\alpha}_{\beta\gamma}$ are the Cristoffel symbols of the second kind \cite{Kochin}, \cite{Dash}
which equal zero identically in any flat space. It is clear that he last term in the action function $S$ is 
not varied and we do not to know the exact numerical value of the constant $a$ in Eq.(\ref{actionu}). Also for 
the non-flat spaces in the last term we have to replace $d\Omega \rightarrow \sqrt{- g} d\Omega$.    

\section{Coulomb's law in multi-dimensions} 

The explicit form of the Coulomb interaction between two point, electrically charged particles is of a 
crucial importance for our present purposes. In Gauss units, which are used almost everywhere in this 
study, the Coulomb's law for three-dimensional space has a very simple form $V(r_{21}) = \frac{q_1 
q_2}{r_{21}}$, where $V(r_{21})$ is the Coulomb potential, $q_1$ and $q_2$ are the electric charges of 
the two point particles (1 and 2) and $r_{21}$ is the interparticle distance which equals $r = 
\sqrt{(x_2 - x_1)^{2} + (y_2 - y_1)^{2} + (z_2 - z_1)^{2}}$, where $(x_1, y_1, z_1)$ and $(x_2, y_2, 
z_2)$ are the Cartesian coordinates of the two interacting particles. Note that the Coulomb interaction 
potential does not contain the factor $4 \pi$. Furthermore, the Coulomb potential essentially coincides 
with the singular part of the Green's function for the three-dimensional Laplace operator, i.e., 
$V(r_{21}) = q_1 q_2 G({\bf r}_1, {\bf r}_2) = q_1 q_2 G(\mid {\bf r}_1 - {\bf r}_2 \mid) = \frac{q_1 
q_2}{\mid {\bf r}_2 - {\bf r}_1 \mid}$ and $\Delta \Bigl(\frac{1}{\mid {\bf r}_2 - {\bf r}_1 
\mid}\Bigr) = \nabla^{2} \Bigl(\frac{1}{\mid {\bf r}_2 - {\bf r}_1 \mid}\Bigr) = \nabla 
\Bigl(\frac{{\bf r}_1 - {\bf r}_2}{\mid {\bf r}_2 - {\bf r}_1 \mid^{3}}\Bigr) = - 4 \pi \delta({\bf 
r}_2 - {\bf r}_1)$. The last equation can also be re-written for the intensity of electric field ${\bf 
E}$, which is the negative gradient of the potential $\varphi$. This equation takes the familiar form 
$div {\bf E} = - \nabla \Bigl[ \nabla \Bigl(\frac{q_1 q_2}{r_{21}}\Bigr) \Bigr] = q_1 q_2 \; \nabla 
\Bigl(\frac{{\bf r}_{21}}{r^{3}_{21}}\Bigr) = 4 \pi \rho({\bf r}_{21})$, where $\rho({\bf x})$ is 
a continuous charge density. The derived expression coincides with the well known differential form 
of Gauss's law of electrostatic and one of the Maxwell equations. These two properties (or two 
criteria) of three-dimensional Coulomb potential will play a crucial role in our definition of the 
multi-dimensional Coulomb potential (see below). 

Now, we need to define the Coulomb potential in multi-dimensional (or $n-$dimensional) space. It is 
a crucial moment for the Maxwell electrodynamics in multi-dimensional spaces which we try to develop 
in this study. Any mistake in such a definition will cost too much for our present purposes. In this 
sense, this Section was a most difficult part of our analysis and it was re-written quite a few times. 
Indeed, we cannot send someone even in the four-dimensional (geometrical) space to repeat the well 
known Coulomb and Cavendish experiments. Therefore, we need to find a way to make an analytical 
generalization of the Coulomb potential to multi-dimensional spaces. In respect to our first criterion
formulated above the Coulomb potential in the $n-$dimensional space must coincide with the singular 
part of the Green's function defined for the multi-dimensional (or $n-$dimensional) Laplace operator 
$\Delta = \Delta_{n} = \frac{\partial^{2}}{\partial x^{2}_{1}} + \frac{\partial^{2}}{\partial x^{2}_{2}} 
+ \ldots + \frac{\partial^{2}}{\partial x^{2}_{n}}$. This leads \cite{Sok1} to the following general 
expression for the Coulomb potential in $n-$dimensional space: $V(r) = b \frac{q_1 q_2}{r^{n-2}_{21}} = 
b \frac{q_1 q_2}{r^{n-2}}$, where $b$ is some numerical factor, $n \ge 3$ and the explicit expression 
for the interparticle distance $r_{21} = r$ takes the multi-dimensional form $r = \sqrt{[x^{(1)}_2 - 
x^{(1)}_1]^{2} + [x^{(2)}_2 - x^{(2)}_1]^{2} + \ldots + [x^{(n)}_2 - x^{(n)}_1]^{2}}$. Here $(x^{(1)}_1, 
x^{(2)}_1, \ldots, x^{(n)}_1)$ and $(x^{(1)}_2, x^{(2)}_2, \ldots, x^{(n)}_2)$ are the Cartesian 
coordinates of the two interacting particles in $n-$dimensional Euclidean space. The $n-$dimensional 
radius $r = \sqrt{[ x^{(1)} ]^2 + [ x^{(2)} ]^{2} + \ldots + [ x^{(n)} ]^2}$ is, in fact, the 
hyper-radius of this point particle. To derive the explicit formula for the Coulomb potential in 
$n-$dimensional space we have applied the method developed by A. Sokolov (see, e.g., \cite{Sok1}, 
\cite{Sok2} and earlier references therein) which allows one to determine the Green's functions for an 
arbitrary linear differential operator.  

In order to determine the factor $b(n)$ we apply the second criterion (see above) which states that 
he Gauss's law must be written in the form $\nabla {\bf E} = f(n) q_1 q_2$, where $f(n)$ is a pure 
angular (or hyper-angular for $n \ge 4$) factor. From here one finds that $b = \frac{1}{n - 2}$ and 
the explicit formula for Coulomb's law in $n-$dimensional space takes the final form $V(r) = 
\frac{q_1 q_2}{(n - 2) r^{n-2}_{21}}$. Now, let us consider a slightly different problem. Suppose, 
we have to determine the static multi-dimensional Coulomb potential $\varphi(r)$ and the corresponding 
intensity of electric field ${\bf E}$ which are generated by a point particle with the electric charge 
$Q$. For this problem, we write the  following formulas for the potential $\varphi$ and for the field 
strength ${\bf E}$: $\varphi = \frac{Q}{(n - 2) r^{n-2}}$ and ${\bf E}  = - \nabla \varphi = \; 
\frac{Q {\bf n_{r}}}{r^{n-1}}$, where ${\bf n}_{r}$ is the unit vector ${\bf n}_{r} = \frac{{\bf 
r}}{r}$ which is directed from the electric charge $Q$ to an observation point. To write the Gauss's 
law in multi-dimensional space let us assume that a point electrical charge $Q$ is located inside (and 
outside) of a closed $(n - 1)$ dimensional hyper-surface. In this case $r$ is the distance from the  
charge to a point on the hyper-surface, ${\bf n}$ is outwardly directed normal ${\bf n} = \frac{{\bf 
r}}{r}$ to the surface at that point and $da$ is the element of the surface area. Then for the normal 
component of ${\bf E}$ times the area element we can write 
\begin{equation}
   ({\bf E} \cdot {\bf n}) da = Q \; \; \frac{cos\Theta}{r^{n-1}} \; \; da = Q \; \; \frac{r^{n-1} 
   d\Omega}{r^{n-1}} =  Q d\Omega \; \; , \; \; \label{GaussA} 
\end{equation}
where $d\Omega$ is the element of solid hyper-angle (in $n-$dimensional space) subtended by $da$ at the 
position of the charge. It is important here that the ${\bf E}$ is directed along the line from the 
hyper-surface element to the charge $Q$. This means that we have found no contradiction here between 
out two criteria and and Eq.(\ref{GaussA}), since the following hyper-angular integration over $\Omega$
does produce only an additional pure hyper-angular factor $f(n)$. 

Now, by integrating the normal component of {\bf E} over the whole hyper-surface, it is easy to find that 
\begin{equation}
 \oint ({\bf E} \cdot {\bf n}) da =  Q \oint d\Omega = Q \; \; \frac{n 
 \pi^{\Bigl(\frac{n}{2}\Bigr)}}{\Gamma\Bigl( 1 + \frac{n}{2} \Bigr)} = f(n) Q \; \; ,  \; \label{GaussB} 
\end{equation}
where $f(n) = \frac{n \pi^{\Bigl(\frac{n}{2}\Bigr)}}{\Gamma\Bigl( 1 + \frac{n}{2} \Bigr)}$ is the 
geometrical (or hyper-angular) factor. In this equation the symbol $\Gamma(x)$ stands for the Euler's 
gamma-function (or Euler's integral of the second kind). It can be shown (see, e.g., \cite{GR}) that 
$\Gamma(1 + x) = x \Gamma(x)$ and $\Gamma\Bigl( \frac12 \Bigr) = \sqrt{\pi}$. The formula, 
Eq.(\ref{GaussB}), is true, if the charge $Q$ lies inside of the $n-$dimensional hyper-surface. However, 
if this charge lies outside of this hyper-surface the expression in the right-hand side of 
Eq.(\ref{GaussB}) equals zero identically. Thus, we have reproduced the Gauss's law in multi-dimensional
spaces for a single point charge $Q$. For a discrete set of point charges and for a continuous charge 
density $\rho({\bf r})$ the Gauss's law becomes:
\begin{equation}
 \oint ({\bf E} \cdot {\bf n}) da = \frac{n \pi^{\Bigl(\frac{n}{2}\Bigr)}}{\Gamma\Bigl( 1 + \frac{n}{2} 
 \Bigr)} \; \; \; \sum^{K}_{k=1} Q_k \; \; = f(n) \; \; \sum^{K}_{k=1} Q_k \; \; \; \label{GaussC}
\end{equation}
and 
\begin{equation}
 \oint ({\bf E} \cdot {\bf n}) da = \frac{n \pi^{\Bigl(\frac{n}{2}\Bigr)}}{\Gamma\Bigl( 1 + \frac{n}{2} 
 \Bigr)} \; \; \; \int_{V} \rho({\bf r}) d^{n}{\bf r} \; \; = f(n) \; \; \int_{V} \rho({\bf r}) 
 d^{n}{\bf r} \; \; \; \label{GaussD}
\end{equation}
respectively. In Eq.(\ref{GaussC}) the sum is over only those charges inside of the hyper-surface $S$, 
while in Eq.(\ref{GaussD}) the volume (or hyper-volume) enclosed by $S$. 

The differential form of these equations in $n-$dimensional Euclidean space is: 
\begin{equation}
  div {\bf E} = - div \; \Bigl( grad \; \varphi \Bigr) = - \Delta \varphi = \frac{n 
  \pi^{\Bigl(\frac{n}{2}\Bigr)}}{\Gamma\Bigl( 1 + \frac{n}{2} \Bigr)} \; \rho({\bf r}) = f(n) \rho({\bf 
  r}) \; \; , \; \label{DivE} 
\end{equation}
where $f(n) = \frac{n \pi^{\Bigl(\frac{n}{2}\Bigr)}}{\Gamma\Bigl( 1 + \frac{n}{2} \Bigr)}$ is the 
geometrical (or hyper-angular) factor which is the volume $V_n$ of $n-$dimensional unit ball times
the dimension $n$ of geometrical space. In other words, the factor $f(n)$ is the surface area $S_n$ 
of $n-$dimensional unit ball, since the equality $S_n = n V_n$ is always obeyed for the $n-$dimensional 
unit ball \cite{Fland} and $n$ is an integer positive number. The physical sense of this factor $f(n)$ 
is simple: it is the total hyper-angle defined for a single point (central) particle located in the 
$n-$dimensional space. For the system of a few discrete charges one has to replace $\rho({\bf r})
\rightarrow \sum^{K}_{k=1} Q_k$, etc. 

The $n-$dimensional hyper-angular factor $f(n)$ from Eq.(\ref{GaussB}) plays a central role in our 
development of Maxwell electrodynamics in multi-dimensional spaces. In particular, the knowledge of 
this factor allows one to write the explicit formula for the action function (or action integral) of 
the electrically charged particles which move in the multi-dimensional (or $n-$dimensional) 
electromagnetic field. This problem is considered below.  

\section{Action function and Maxwell equations in multi-dimensional flat spaces} 

In this Section we consider the Maxwell's equation in multi-dimensional flat spaces, e.g., in pseudo-Euclidean 
spaces. Results derived below will extensively be used in the following Sections of this study.  First of all,
by using the factor $f(n)$ obtained in Eq.(\ref{GaussB}) we can write the final expression for the action 
function $S$ in Gauss units
\begin{eqnarray}
 S = S_p + S_{fp} + S_f = - \sum_k \int m_k c ds - \frac{1}{c^{2}} \int A_{\alpha} j^{\alpha} dx^{\alpha} 
 - \frac{1}{4 c f(n)} \int F_{\alpha\beta} F^{\alpha\beta} d\Omega \; \; , \; \label{action} 
\end{eqnarray}
where $\frac14$ (or $- \frac14$) is the Heaviside constant, $c$ is the speed of light in vacuum, while 
$j^{\alpha}$ is the electric current (or simply, current) in $(n + 1)-$dimensional space. By varying all 
components of the $\bar{A}$ vector in this action integral, Eq.(\ref{action}), we derive the second 
group of Maxwell's equations, Eq.(\ref{Maxweq2}), which contains, in the general case, the non-homogeneous 
differential equations. By omitting some obvious details we can write the complete set of Maxwell's 
equations in the following tensor form 
\begin{eqnarray}
 \frac{\partial F_{\gamma\lambda}}{\partial x^{\beta}} + \frac{\partial F_{\lambda\beta}}{\partial 
  x^{\gamma}} + \frac{\partial F_{\beta\gamma}}{\partial x^{\lambda}} = 0 \; \; \; \label{Maxweq1}
\end{eqnarray} 
and 
\begin{eqnarray}
 \frac{\partial F^{\alpha\beta}}{\partial x^{\beta}} = - \frac{n \pi^{\Bigl(\frac{n}{2}\Bigr)}}{c 
 \Gamma\Bigl( 1 + \frac{n}{2} \Bigr)} \; j^{\alpha} = - \frac{f(n)}{c} j^{\alpha} \; \; , \; 
 \label{Maxweq2} 
\end{eqnarray}
where $j^{\alpha}$ is the $(n + 1)-$dimensional current-vector (or current, for short) defined above. All 
equations from the both groups of these equations, Eqs.(\ref{Maxweq1}) - (\ref{Maxweq2}), are the first-order 
differential equations upon spatial coordinates and time $t$ (or temporal coordinate). From Eq.(\ref{Maxweq1}) 
one finds the following condition for the current 
\begin{eqnarray}
 \frac{\partial^{2} F^{\alpha\beta}}{\partial x^{\alpha} \partial x^{\beta}} = - \frac{f(n)}{c} 
 \frac{\partial j^{\alpha}}{\partial x^{\alpha}} = 0 \; \; \; . \; \; \label{Maxweq2} 
\end{eqnarray}
This result is obvious, since application of any symmetric operator (upon $\alpha \leftrightarrow \beta$ 
permutation), e.g., the $\frac{\partial^{2}}{\partial x^{\alpha} \partial x^{\beta}}$ operator, to the truly 
antysimmetric $F^{\alpha\beta}$ tensor always gives zero. Thus, the equality $\frac{\partial 
j^{\alpha}}{\partial x^{\alpha}} = 0$ derived here is a necessary condition for any actual electric current. 
Note also that this equation is written in the form of $(n + 1)-$dimensional divergence. In respect to the 
second Noether's theorem this equation $\frac{\partial j^{\alpha}}{\partial x^{\alpha}} = 0$ means some 
conservation law. It is easy to understand that this law describes conservation of the total electric charge.  

A very close similarity between Maxwell equations derived for multi-dimensional spaces, where $n \ge 3$, and 
analogous Maxwell equations known in three-dimensional space is an obvious fact. However, in some cases this 
leads to fundamental mistakes and most of such mistakes are originated by Eq.(\ref{Maxweq1}). Note here that 
in $n-$dimensional geometrical space we have exactly $n$ components of the intensity of electric field ${\bf 
E}$ and $\frac{n (n + 1)}{2}$ intensities of magnetic field ${\bf H}$. For $n = 3$ (and only in this case) 
we have equal numbers of components in the both ${\bf E}$ and ${\bf H}$ vectors. This leads to the well-known 
vector form of Maxwell electrodynamics. However, already for $n = 4$ the electric field has four components, 
while the magnetic field has six components. When $n$ increases, then the total number of components of the 
magnetic field grow rapidly (quadratically) and significantly exceeds the analogous number of components of 
the electric field. This fact substantially complicates derivation of Maxwell equations written in terms of 
the intensities of electric and magnetic fields in multi-dimensional spaces. Plus, we have a certain problem 
with general definition of the $curl$ (or $rot$) operator in such cases.  

Another interesting result follows from the analysis of tensor equations, Eq.(\ref{Maxweq1}). If one of the 
indexes in this equation equal zero, then this group of equations gives us Faraday's law in multi-dimensional 
space which describes the time-evolution of the magnetic field and it is written in the form of $n$ equations. 
This is good, but what is about other $\frac{n (n - 1) (n - 2)}{6}$ equations which are also included in tensor 
equations Eq.(\ref{Maxweq1})? After some transformations one finds that these additional equations are written 
in the form where three-dimensional divergences of some three-dimensional pure-magnetic vectors equal zero. By 
the pure magnetic vectors we mean vectors assembled from the space-like components of the field tensor $F^{pq}$ 
(or $F_{pq}$) only (for flat spaces it is always possible). Based on ideas by Dirac \cite{Dir48} we can 
formulate this result in the following form: {\it the magnetic field cannot have sources neither in our 
three-dimensional space, nor in any three-dimensional subspace of multi-dimensional spaces}. This fundamental 
statement is directly and very closely related to the discrete nature of electric charge. Furthermore, the 
correctness of Maxwell electrodynamics (in any space) is essentially based on this statement. By taking into 
account arguments from \cite{Amaldi} we can re-formulate this our statement in the form: {\it The existence of 
magnetic monopoles in our three-dimensional space and, in general, in any three-dimensional subspace of 
multi-dimensional spaces, is strictly prohibited}. Otherwise, the Maxwell electrodynamics will not be correct
and must be replaced by a different approach.  

To conclude this Section, let us present the explicit formula for the energy momentum tensor in multi-dimensional 
space. Definition of this tensor and all details of its calculations are well described in \cite{LLTF}. Therefore, 
here we can only present a few basic formulas, which will be used below in Section VI. The explicit formula for 
the non-symmetrized energy momentum tensor is 
\begin{eqnarray}
 T_{\alpha}^{\beta} = \frac{1}{f(n)} \Bigl( \frac{\partial A_{\gamma}}{\partial x_{\alpha}} F^{\gamma\beta} + 
 \frac14 g_{\alpha}^{\beta} F_{\gamma\rho} F^{\gamma\rho} \Bigr) \; \; , \; \; \label{tensrM}
\end{eqnarray}
where the factor $f(n)$ is the hyper-angular (or geometrical) factor mentioned above. After symmetrization this 
tensor takes the form
\begin{eqnarray}
 T_{\alpha}^{\beta} = \frac{1}{f(n)} \Bigl( F_{\alpha\gamma} F^{\beta\gamma} + \frac14 g_{\alpha}^{\beta} 
 F_{\gamma\rho} F^{\gamma\rho} \Bigr) \; \; , \label{tensrM}
\end{eqnarray}
where $g_{\alpha}^{\beta} = \delta_{\alpha}^{\beta}$ is the substitution tensor \cite{Kochin}. The corresponding 
co- and contravariant tensors are: 
\begin{eqnarray}
  T_{\alpha\beta} = \frac{1}{f(n)} \Bigl( F_{\alpha\gamma} F^{\gamma}_{\beta} + \frac14 g_{\alpha\beta} 
  F_{\gamma\rho} F^{\gamma\rho} \Bigr) \; \; \; {\rm and} \; \; \; T^{\alpha\beta} = \frac{1}{f(n)} 
  \Bigl(  g^{\alpha}_{\gamma} F_{\beta\gamma} F^{\beta\gamma} + \frac14 g^{\alpha\beta} F_{\gamma\rho} 
  F^{\gamma\rho} \Bigr) \; \; , \label{tensrCC} 
\end{eqnarray}
where $f(n) = \frac{n \pi^{\frac{n}{2}}}{\Gamma\Bigl( 1 + \frac{n}{2} \Bigr)}$ is the geometrical (or 
hyper-angular) factor. 

\section{Hamiltonian of the electromagnetic field in multi-dimensional flat spaces} 

The second goal of this study is to develop the Hamiltonian formulation of the multi-dimensional electrodynamics. 
First, let us obtain the explicit formula for the Hamiltonian $H$ of the free electromagnetic field in 
multi-dimensional flat spaces. By using the formula, Eq.(\ref{action}), for the action integral we can write the 
Lagrangian $L$ of the free electromagnetic field in multi-dimensional pseudo-Euclidean space (in Heaviside units)
\begin{eqnarray}
  L = - \frac14 \int F_{\alpha\beta} F^{\alpha\beta} \; d^{n}{\bf x} = - \frac14 \int F^{\alpha\beta} F_{\alpha\beta} 
 \; d^{n}{\bf x} \; \; \; , \; \label{Lagr}
\end{eqnarray}
where $F_{\mu\nu} = A_{\nu,\mu} - A_{\mu,\nu}$ is the electromagnetic field tensor which is antisymmetric $F_{\mu\nu} 
= - F_{\nu\mu}$. From here one finds the following equality $A_{\mu,\nu} = - F_{\mu\nu} + A_{\nu,\mu} = F_{\nu\mu} + 
A_{\nu,\mu}$. Variations of this Lagrangian are written in the following general form 
\begin{eqnarray}
 \delta L = - \frac12 \int F_{\alpha\beta} \delta F^{\alpha\beta} d^{n}{\bf x} = - \frac12 \int F^{\alpha\beta} 
 \delta F_{\alpha\beta} d^{n}{\bf x} \; , \; \label{LagrA}
\end{eqnarray}
where $d^{n}{\bf x}$ means $dx^{1} dx^{2} \ldots dx^{n}$ and the integration is over $n-$dimensional space. Note that 
all integrals considered in this Section are the spatial integrals which contain no integration over the temporal (or 
time) variable. Furthermore, in this Section we shall apply only the Heaviside units. The use of Gauss units 
complicates all formulas below, including the expressions for the momenta. 

In order to develop the Hamiltonian approach for the electromagnetic field we need to consider all variations of the 
velocities for each component of the $(n + 1)$-dimensional vector potential $\bar{A}$. In other words, below we deal 
with variations of the $A_{\mu,0}$ derivatives only, where $\mu$ = 0, 1, $\ldots, n$. In other words, in our 
Hamiltonian formulation all components of the $(n + 1)$-dimensional vector potential $\bar{A}$, i.e., $A_0, A_1, 
\ldots, A_{n}$ components, are the generalized coordinates of our problem. For variations of the velocities 
$A_{\mu,0}$ our formula, Eq.(\ref{LagrA}), for $\delta L$ is written in the form
\begin{eqnarray}
 \delta L = \int F^{\alpha 0} \delta A_{\alpha,0} \; d^{n}{\bf x} = \int B^{\alpha} \delta A_{\alpha,0} \; 
 d^{n}{\bf x} \; , \; \label{LagrA1}
\end{eqnarray}
where $B^{\alpha} = F^{\alpha 0}$ are the contravariant components of the $(n + 1)-$dimensional vector momenta 
$\bar{B}$. In fact, this equation must be considered as the explicit definition of momenta. However, from this 
definition and antisymmetry of the electromagnetic field tensor one finds $B^{0} = F^{00} = 0$. This means that 
0-component of momenta $\bar{B}$ of the electromagnetic field, i.e., $B^{0}$, must be equal zero at all times. 
According to Dirac \cite{Dir64} all similar equations derived at this stage of the Hamiltonian procedure are the 
primary constraints. In our current case this constraint is better to write in the form of a weak identity $B^{0} 
\approx 0$.  

By using our momenta $B^{\alpha}$ we can introduce the Hamiltonian of the free electromagnetic field in 
multi-dimensional pseudo-Euclidean (flat) space
\begin{eqnarray}
 H &=& \int B^{\alpha} A_{\alpha,0} \; d^{n}{\bf x} - L = \int \Bigl( F^{q 0} A_{q,0} + \frac14 F^{p q} F_{p q} 
 + \frac14 F^{p 0} F_{p 0} + \frac14 F^{0 p} F_{0 p} \Bigr) d^{n}{\bf x} \nonumber \\
  &=& \int \Bigl( F^{q 0} A_{q,0} + \frac14 F^{p q} F_{p q} + \frac12 F^{p 0} F_{p 0} \Bigr) d^{n}{\bf x} = 
 \int {\cal H} d^{n}{\bf x} \; \; , \; \label{HamltA1}
\end{eqnarray}
where ${\cal H}$ is the Hamiltonian space-like density (scalar) which is 
\begin{eqnarray}
 {\cal H} = F^{q 0} A_{q,0} + \frac14 F^{p q} F_{p q} + \frac12 F^{p 0} F_{p 0} \; \; \label{Hamltden}
\end{eqnarray}
For the $A_{q,0}$ derivative we substitute its equivalent expression $A_{q,0} = - F_{q 0} + A_{0,q}$ (see above) 
and obtain
\begin{eqnarray}
 H = \int \Bigl( \frac14 F^{p q} F_{p q} - \frac12 F^{p 0} F_{p 0} + F^{q 0} A_{0,q} \Bigr) d^{n}{\bf x} =
 \int  \Bigl( \frac14 F^{p q} F_{p q} + \frac12 B^{p} B^{p} + B^{q} A_{0,q} \Bigr) d^{n}{\bf x} \; \; . \; 
 \label{HamltA2} 
\end{eqnarray}

In the last term of this Hamiltonian we can do a partial integration which actually leads to the following 
replacement $F^{q 0} A_{0,q} \rightarrow - A_{0} \frac{\partial F^{q 0}}{\partial x_{q}} = - A_{0} 
( B^{q} )_q = - A_{0} B^{p}_{,p}$. This reduces our Hamiltonian, Eq.(\ref{HamltA2}), to the form
\begin{eqnarray}
 H = \int \Bigl( \frac14 F^{p q} F_{p q} - \frac12 F^{p 0} F_{p 0} + F^{q 0} A_{0,q} \Bigr) d^{n}{\bf x} = 
 \int \Bigl( \frac14 F^{p q} F_{p q} + \frac12 B^{p} B^{p} -  A_{0} B^{p}_{,p} \Bigr) d^{n}{\bf x} \; \; 
 . \; \label{HamltA3}
\end{eqnarray}
This is the Hamiltonian of the free electromagnetic field written in the closed analytical form. The 
corresponding Hamiltonian space-like density takes the form
\begin{eqnarray}
  {\cal H} = \frac14 F^{p q} F_{p q} + \frac12 B^{p} B^{p} -  A_{0} B^{p}_{,p} \; \; . \; \label{Hamltden1}
\end{eqnarray}
Note that by performing these transformations and deriving the Hamiltonian, Eq.(\ref{HamltA3}), we have gained even 
more than we wanted at the beginning of our procedure. In fact, development of any Hamiltonian approach means that 
we have a simplectic structure which is defined by the Poisson brackets between basic dynamical (Hamiltonian) 
variables: $(n + 1)$ coordinates $A_{\mu}$ and $(n + 1)$ momenta $B^{\mu}$. These Poisson brackets are 
defined as follows 
\begin{eqnarray}
 [ A_{\mu}(\bar{x}_1), B^{\nu}(\bar{x}_2) ] = g^{\nu}_{\mu} \delta^{(n)}({\bf x}_1 - {\bf x}_2) \; , \; [ 
 A_{\mu}({\bf x}_1), A_{\nu}({\bf x}_2) ] = 0 \; \; , \; [ B^{\mu}({\bf x}_1), B^{\nu}({\bf x}_2) ] = 0 \; , 
 \; \label{PB}  
\end{eqnarray}
where $g^{\nu}_{\mu} = \delta^{\nu}_{\mu}$ is the Kronecker delta-function, while $\mu = 0, 1, \ldots, n$ and $\nu 
= 0, 1, \ldots, n$. 

In general, the Poisson brackets are used as the main working tool in any Hamiltonian approach developed for a given 
physical system. Moreover, these brackets allow one to introduce a simplectic $(2 n + 2)-$dimensional phase space of 
the Hamiltonian variables $\{ A_{\alpha}, B^{\beta} \}$ which are defined in each point $\bar{x}$ of the $(n + 
1)-$dimensional space-time manifold. The original configuration space of this problem is the direct sum of the $(n + 
1)-$dimensional subspace of $A_{\mu}-$coordinates and $(n + 1)-$dimensional subspace of $A_{\mu,0}-$velocities. In 
turn, this allows one to consider and apply various canonical transformations of the Hamiltonian canonical variables. 
Furthermore, by using the Poisson brackets Eq.(\ref{PB}) we can complete our Hamiltonian approach for the classical 
electrodynamics and perform its quantization. 

To illustrate this fact let us go back to the primary constraint $B^{0} \approx 0$ mentioned above. This constraint 
must remain satisfied at all times. This means its time derivative $\frac{d B^{0}}{dt}$, which in our Hamiltonian 
approach equals to the Poisson bracket $[ B^{0}, H ]$, must be zero at all times. This Poisson bracket is easily 
determined, since in the Hamiltonian, Eq.(\ref{HamltA3}), there is only one term (the last term) which does not 
commute with the momentum (or primary constraint) $B^{0}$ 
\begin{eqnarray}
 [ B^{0}, \frac14 F^{p q} F_{p q} - \frac12 F^{p 0} F_{p 0} - A_{0} B^{q}_{q} ] = - [ B^{0}, A_{0} ] B^{p}_{,p} 
 = [ A_0, B^{0}] B^{p}_{,p} = B^{p}_{,p} \label{second} 
\end{eqnarray}
In other words, we have found another weak equality $B^{p}_{,p} \approx 0$ which must be obeyed at all times. 
According to Dirac \cite{Dir50} and \cite{Dir64} this condition is the secondary constraint of our Hamiltonian 
formulation of multi-dimensional Maxwell theory of radiation. The next Poisson bracket $[ B^{p}_{,p}, H ]$ (or 
$[ B^{p}_{,p}, {\cal H} ]$) equals zero identically, which indicates clearly that the chain of first-class 
constraints is closed and our Hamiltonian formulation does not lead to any tertiary and/or other constraints 
of higher order. Briefly, this means the complete closure (or Dirac closure) of the Hamiltonian procedure for 
the free electromagnetic field in multi-dimensional space.    

\subsection{Further transformations of the Hamiltonian}

The first term in the Hamiltonian of the free electromagnetic field in multi-dimensional space, Eq.(\ref{HamltA3}), 
includes a number of different terms, but it does not contain any of the canonical variables. It is hard to use 
such a Hamiltonian for analysis and solution of actual problems in classical and/or quantum quantum electrodynamics. 
Therefore, we have to transform this Hamiltonian to the form which explicitly contain canonical variables in each 
term. Then, our newly derived Hamiltonian $H$ and/or the corresponding Hamiltonian density ${\cal H}$ can be applied 
for solution of many actual problems. For convenience, below we shall deal with the Hamiltonian density ${\cal H}$. 
Partial integration of the first term in the Hamiltonian, Eq.(\ref{HamltA3}), leads to the following expression for 
the Hamiltonian density Eq.(\ref{Hamltden1}):    
\begin{eqnarray}
 {\cal H} = \Bigl(F^{q p}\Bigr)_{q} A_{p} + \frac12 B^{p} B^{p} -  A_{0} B^{p}_{,p} = \Bigl( \frac{\partial^{2} 
 A^{p}}{\partial x_{q} \partial x^{q}} - \frac{\partial^{2} A^{q}}{\partial x_{q} \partial x_{p}} \Bigr) A_{p} 
 + \frac12 B^{p} B^{p} - A_{0} B^{p}_{,p} \; , \; \label{Hamltden3}
\end{eqnarray}
where $p = 1, 2, \ldots, n$ and $q = 1, 2, \ldots, n$. For this Hamiltonian density we can write the following 
system of canonical equations 
\begin{eqnarray}
 \frac{d A_{p}}{d t} = [ A_{p}, {\cal H} ] = \frac12 \; (2 B^{p}) = B^{p} \; \; \; \label{caneq1}
\end{eqnarray}
and 
\begin{eqnarray}
 \frac{d B^{p}}{d t} = [ B^{p}, {\cal H} ] = - \Bigl( \frac{\partial^{2} A^{p}}{\partial x_{q} \partial x^{q}} 
 - \frac{\partial^{2} A^{q}}{\partial x_{q} \partial x_{p}} \Bigr) = \frac{\partial^{2} A_{p}}{\partial x_{q} 
 \partial x^{q}} - \frac{\partial^{2} A_{q}}{\partial x_{q} \partial x_{p}} \; \; . \; \label{caneq2}
\end{eqnarray}
Combining these two equations one finds 
\begin{eqnarray}
 \frac{d^{2} A_{p}}{d t^{2}} = \frac{\partial^{2} A_{p}}{\partial x_{q} \partial x^{q}} - \frac{\partial^{2} 
 A_{q}}{\partial x_{q} \partial x_{p}} \; \; . \; \label{eqmot2}
\end{eqnarray}
Taking into account the gauge condition $\frac{\partial A_{q}}{\partial x_{q}} = 0$ (see below) we reduce the 
last equation to the form 
\begin{eqnarray}
 \frac{\partial^{2} A_{p}}{\partial t^{2}} - \frac{\partial^{2} A_{p}}{\partial x_{q} \partial x^{q}} = 0 \; 
 \; , \; {\rm or} \; \; \frac{\partial^{2} {\bf A}}{\partial t^{2}} - \Delta {\bf A} = 0 \; , \; \label{caneq2}
\end{eqnarray}
which is the wave equations written in the $(n + 1)-$dimensional space-time. The $n-$dimensional Laplace 
operator $\Delta$ in this equation is 
\begin{eqnarray}
 \Delta = \frac{\partial^{2} }{\partial x_{q} \partial x^{q}} = g^{qr} \frac{\partial^{2} }{\partial x^{q} 
 \partial x^{r}} = g_{qr} \frac{\partial^{2} }{\partial x_{q} \partial x_{r}} \; \; . \; \label{Lapl}
\end{eqnarray} 

Thus, in our Hamiltonian approach the multi-dimensional wave equation for the free electromagnetic field is 
derived as a direct consequence of the canonical Hamilton equations obtained for this field. Such a derivation 
of the wave equation for a free electromagnetic field described here is, probably, the most direct, fast, and 
logically clear of all known (alternative) methods. In addition to this, we have rigorously derived the two 
additional conditions for the momenta of the free electromagnetic filed: $B^{0} \approx 0$ and $B^{p}_{,p} 
\approx 0$. In our Hamiltonian formulation these two weak equations are called the primary and secondary 
constraints, respectively. It is easy to show that these two constraints are first-class \cite{Dir64}. In the 
four-dimensional case Dirac has suggested \cite{Dir64} that these two constraints are the generators (or 
generating functions) for infinitesimal contact transformations which do not change the actual physical state 
of the free electromagnetic field, i.e., they are two independent generators of internal symmetry. Twenty 
years later this statement has rigorously been proven by L. Castellani \cite{Cast}. All these results are the 
great and obvious advantages of the Dirac's (Hamiltonian) formulation of the Maxwell theory. Now, by using all 
first-class constraints, which have been derived during the Hamiltonian formulation, one can determine the true 
symmetry of any given physical field. For the free electromagnetic field such a symmetry group coincides with 
the Lorentz $SO(3,1)$-group. In general, by operating with the first-class constraints only it is impossible to 
restore the so-called hidden (or additional) symmetries of the free electromagnetic field. For instance, for 
the free electromagnetic field considered in three-dimensional space the complete group of point symmetry is 
the $SO(4,2)$-group which has fifteen generators \cite{BHag}, while the Lorentz $SO(3,1)$-group has only six 
generators. The powerful method of Bessel-Hagen \cite{BHag} is based on applications of the second Noether's 
theorems which is applied to the Lagrangian of the free electromagnetic field. In this short paper we cannot 
discuss all details of this interesting problem.  

\subsection{First-class constraints and gauge invariance} 

In this Section we consider a different symmetry (or invariance) of Maxwell equations which is directly and 
closely related to the primary and secondary first-class constraints. This invariance is the well known gauge 
invariance (or symmetry) of the Maxwell equations. The gauge invariance of three-dimensional Maxwell equations 
has been studied by many famous authors, including Heitler \cite{Heitl}, Jackson \cite{Jack}, \cite{Jack1}, 
Gelfand and Fomin \cite{GF} and others (see, e.g., \cite{Wol}). Briefly, the gauge invariance means that we 
can impose some additional conditions upon the physical fields, or some of their components, and these 
additional conditions do not change solutions of the original problem (but they can change equations!). The 
gauge conditions are often used to simplify the Hamiltonian equations of motion either by reducing the total 
number of variable fields, or by vanishing some terms (or combinations of terms) in these equations. Let us 
discuss the gauge invariance of the free electromagnetic field (or 'pure radiation field' \cite{Heitl}) by 
using the two first-class constraints which we have derived above: $B^{0} \approx 0$ and $B^{p}_{,p} \approx 
0$. By re-writing these two constraints in terms of the components of the $(d + 1)-$dimensional vector 
potential $\bar{A} = (\varphi, {\bf A})$ and their temporal derivatives, one finds 
\begin{eqnarray}
 B^{0} \approx 0 \Rightarrow \frac{\partial \varphi}{\partial t} = 0 \; \; {\rm and} \; \; B^{p}_{,p} 
 \approx 0 \Rightarrow \frac{\partial}{\partial t} \Bigl( div {\bf A} \Bigr) = 0 \; \; \label{phi-A}
\end{eqnarray}
where we used the traditional sign of actual equality '=' instead of the weak equality '$\approx$', which 
has been used above in the Dirac's Hamiltonian approach. The two equalities in the right-hand side of 
Eq.(\ref{phi-A}) lead us to the two following to the following equations: $\varphi = \varphi({\bf r})$ and 
$div {\bf A} = C({\bf r})$, where the scalars $\varphi({\bf r})$ and $C({\bf r})$ are the functions of $n$ 
spatial coordinates only, and they do not change with time, i.e., they are time-independent scalar functions. 
It is clear that these two time-independent scalars are not related in any way to the Hamiltonian formulation 
of the Maxwell theory of electromagnetic fields. Indeed, the Hamiltonian approaches describe only 
time-evolution of the Hamiltonian dynamical variables. For static problems there are other different methods. 
Therefore, without loss of generality, we can assume that these time-independent scalars $\varphi({\bf r})$ 
and $C({\bf r})$ equal zero identically at all times. 

Based on these arguments we can write the four following equations for the field dynamical variables (or 
Hamiltonian variables):
\begin{eqnarray}
 \varphi = 0 \; \; , \; \; \frac{\partial \varphi}{\partial t} = 0 \; \; , \; \; div {\bf A} = 0 \; \; 
 {\rm and} \; \; \frac{\partial}{\partial t} \Bigl( div {\bf A} \Bigr) = 0 \; \; , \; \label{phi-Ag}
\end{eqnarray} 
which can be considered as the four independent 'basis vectors'. In general, the set of $N_g$ gauge conditions 
$\psi_{i}$ is represented as a linear combinations of the four basis vectors from Eq.(\ref{phi-Ag}): 
\begin{eqnarray}
 \psi_{i} = \alpha_i \varphi + \beta_{i} \; \frac{\partial \varphi}{\partial t} + \gamma_{i} \; div {\bf A} 
 + \delta_{i} \; \frac{\partial}{\partial t} \Bigl( div {\bf A} \Bigr) = 0 \; \; , \; \label{phi-AgA}
\end{eqnarray} 
where $i$ = 1, 2, 3, 4, while $\alpha_i, \beta_i, \gamma_i$ and $\delta_i$ are some numerical constants. 
Let us discuss the principal question about the number $N_g$ which is the number of sufficient (or essential) 
gauge equations. For the free electromagnetic field $N_g$ equals two, since exactly this number of conditions 
has been found in the Hamiltonian formulation of electrodynamics developed by Dirac (see above). The two 
equations $\frac{\partial \varphi}{\partial t} = 0$ and $\frac{\partial}{\partial t} \Bigl( div {\bf A} \Bigr) 
= 0$ define the so-called Dirac gauge which is discussed above. Formally, for the Dirac gauge we can introduce 
the third gauge condition $\varphi = 0$ and completely exclude the pair of variables $(\varphi, \frac{\partial 
\varphi}{\partial t} \Bigr)$ from the list of our dynamical variables. However, this follows not from some 
general principle, but from the explicit form of the Dirac's Hamiltonian density, Eq.(\ref{Hamltden1}), 
for the pure radiation field (see above), where the only term which includes the scalar potential $\varphi$ is 
written as a product of $\varphi$ (or $A_0$) and secondary constraint $B^{p}_{,p}$. This term equals zero on 
shell of the first-class constraints. 

An alternative choice of two gauge equations $\frac{\partial \varphi}{\partial t} = 0$ and $div {\bf A} = 0$ 
corresponds to the famous Coulomb gauge, which provides the best choice for many three-dimensional QED 
problems in atomic and molecular physics. In the Coulomb gauge the scalar potential $\varphi (= A_{0})$ is 
always a static potential, while the $n-$dimensional vector potential ${\bf A}$ is always transverse. The 
Coulomb gauge and other gauges discussed here are easily generalized for $n-$dimensional spaces. Another 
choice of the basic gauge equations defines the Lorentz gauge. Formally, this gauge is defined by one 
(Fermi's) equation $\frac{\partial \varphi}{\partial t} + div {\bf A} = 0$. In respect to the Dirac theory 
this set of gauge conditions is not complete and a second gauge equation can be added. For instance, one can 
chose the second condition in the form $\frac{\partial \varphi}{\partial t} - div {\bf A} = 0$, which is a 
relativistic invariant for the electromagnetic wave which propagates from the present to the past. A different 
choice of the second equation for the Lorentz gauge corresponds to the so-called Heitler's gauge, which is 
based on the two equations $\frac{\partial \varphi}{\partial t} + div {\bf A} = 0$ and $\frac{\partial}{\partial 
t} \Bigl( div {\bf A} \Bigr) = 0$ for the free electromagnetic field \cite{Heitl}. The advantage of this useful 
gauge is obvious: if these equations hold at $t = 0$, then the equation $\frac{\partial \varphi}{\partial t} + 
div {\bf A} = 0$ is always satisfied. These simple examples of different gauges are mentioned here only to 
illustrate an ultimate power of Dirac's approach which simplifies internal analysis of various gauges. 

Let us discuss the general source of gauges which often arise in different field theories, e.g., in Maxwell 
theory of radiation, metric gravity, tetrad gravity, etc. Here we want to investigate this problem from the 
Hamiltonian point of view. First, let us assume that we have imposed all four conditions from Eq.(\ref{phi-Ag}) 
on our dynamical variables. What does it mean for these variables? The first two equations $\varphi = 0$ 
and $\frac{\partial \varphi}{\partial t} = 0$ mean that the variable $\varphi$ and corresponding momentum $B^{0}$ 
(or velocity $\frac{\partial \varphi}{\partial t})$ are not dynamical (Lagrange) variables of our problem. In 
other words, we have to exclude these two variables before application of our Hamiltonian procedure. The same 
statement is true about the two equations $div {\bf A} = 0$ and $\frac{\partial}{\partial t} \Bigl( div {\bf 
A} \Bigr) = 0$, but $div {\bf A}$ is not a regular dynamical variable of the original problem. In reality, the 
function $div {\bf A}$ appears in the secondary constraint in the Dirac's Hamiltonian formulation developed for 
the pure radiation filed. This function is a linear combination of the first-order derivatives of covariant 
components of the multi-dimensional vector potential ${\bf A}$. The Hamiltonian canonical variables do not 
include any sum of the space-like derivatives of this potential. Therefore, it is not clear how we can exclude 
the scalar $div {\bf A}$ and its time-derivative from the list of our canonical variables. However, the main 
obstacle on the way of exclusion the four variables, Eq.(\ref{phi-Ag}), follows from the fact that we have only 
two gauge equations (not four!). This means that we cannot correctly exclude all four variables and have to keep 
them in our procedure. These 'extra' variables survive our Hamiltonian procedure only in the form of additional 
equations for the Hamiltonian dynamical variables. In other words, the gauge conditions are the integral parts of 
any Hamiltonian approach developed for an arbitrary physical field. This is the general principle which explains 
why different field theories with the first-class constraints always have some number of non-trivial gauge 
conditions (or equations). 

However, this is not the end of the story. Let us look at the constraints in multi-dimensional electrodynamics 
from a different point of view. Consider the following two-parametric $(\alpha,\beta)-$family of the 
Hamiltonian densities 
\begin{eqnarray}
 {\cal H}(\alpha,\beta) = \frac14 F^{p q} F_{p q} + \frac12 B^{p} B^{p} - A_{0} B^{p}_{,p} + \Bigl( \alpha 
 B^{0} + \beta B^{p}_{,p} \Bigr)^{2} \; \; . \; \label{Hamltden3}
\end{eqnarray}
where $B^{0}$ and $B^{p}_{,p}$ are the functions of the canonical variables of the problem. At this moment 
we cannot assume that there are some restrictions on these two quantities. In other words, for now the 
$B^{0}$ and $B^{p}_{,p}$ values are not the constraints yet. 

In general, to operate with the two-parametric family of Hamiltonian densities ${\cal H}(\alpha,\beta)$ 
in some constructive way we have to formulate the following variatonal principle: the actual (or true) 
Hamiltonian density coincides with the minimal Hamiltonian density ${\cal H}(\alpha,\beta)$, 
Eq.(\ref{Hamltden3}), in respect to possible variations of the two numerical parameters $\alpha$ and 
$\beta$. This principle immediately leads to the two following weak identities: 
\begin{eqnarray}
 \Bigl( \alpha B^{0} + \beta  B^{p}_{,p} \Bigr) B^{0} \approx 0 \; \; \; {\rm and} \; \; \; \Bigl( \alpha 
 B^{0} + \beta B^{p}_{,p} \Bigr) B^{p}_{,p} \approx 0  \; \; . \; \label{systemA} 
\end{eqnarray}
One obvious solution of this system gives us the two Dirac's constraints $B^{0} \approx 0$ and $B^{p}_{,p} 
\approx 0$ which have been derived above. In general, there are other solutions of the system 
Eq.(\ref{systemA}), and one of them can be written in the form 
\begin{eqnarray}
  \alpha_1 B^{0} + \beta_1  B^{p}_{,p} \approx 0 \; \; \; {\rm and} \; \; \; \alpha_2 B^{0} + 
  \beta_2 B^{p}_{,p} \approx 0  \; \; . \; \label{systemAA} 
\end{eqnarray}
where the coefficients $\alpha_1, \beta_1, \alpha_2$ and  $\beta_2$ form a regular (i.e., invertible) 
$2 \times 2$ matrix. The principle formulated above is called the optimal principle for the constrained 
motions, since in actual physical systems the motion along first-class constraints is optimal, or it 
can be considered as optimal.  
    
\section{Multi-dimensional Maxwell equations in non-flat spaces}\label{section5}

The Maxwell equations can be written in the covariant form which is more appropriate in applications to the 
metric gravity (or general relativity) in multi-dimensional Riemannian spaces. In this and next Sections we
deal with the multi-dimensional Riemannian spaces only. These spaces are not flat, and they are often called 
the spaces of non-zero curvature. Indeed, the corresponding equations, Eq.(\ref{Maxweq1}) and (\ref{Maxweq2}), 
for the flat multi-dimensional spaces have already been written in the tensor (or covariant) form. Furthermore, 
the electromagnetic field tensor $F_{\alpha\beta}$, which has been defined by Eq.(\ref{tensor}), is truly 
skew-symmetric in respect to permutations of its indexes, i.e., $F_{\alpha\beta} = - F_{\beta\alpha}$ and 
$F^{\alpha\beta} = - F^{\beta\alpha}$. These two facts simplify the process of derivation of the Maxwell 
equations in the covariant form. In fact, to derive the covariant form of Maxwell equations one needs to 
replace all usual derivatives written in Cartesian coordinates by the tensor derivatives. After such a 
replacement the first group of Maxwell equations in multi-dimensional Riemannian spaces takes the form 
\begin{eqnarray}
 \nabla_{\beta} F_{\gamma\lambda} + \nabla_{\lambda} F_{\beta\gamma} + \nabla_{\gamma} F_{\lambda\beta} 
  = 0 \; \; \; ( {\rm or} \; \;  \nabla_{\beta} F_{\gamma\lambda} = \nabla_{\gamma} F_{\beta\lambda} - 
  \nabla_{\lambda}  F_{\beta\gamma} ) \; , \; \label{Maxweq11}
\end{eqnarray}
where $\nabla_{\beta}$ is the tensor (or covariant) derivative, i.e., 
\begin{eqnarray}
 \nabla_{\beta} F_{\gamma\lambda} = \frac{\partial F_{\gamma\lambda}}{\partial x^{\beta}} - 
 \Gamma^{\mu}_{\gamma\beta} F_{\mu\lambda} - \Gamma^{\mu}_{\lambda\beta} F_{\gamma\mu} \; \; 
 \label{Maxweq11a}
\end{eqnarray}
where $\Gamma^{\gamma}_{\alpha\beta} = \frac12 \Bigl( \frac{\partial g_{\gamma\beta}}{\partial x^{\alpha}} + 
\frac{\partial g_{\alpha\gamma}}{\partial x^{\gamma}} - \frac{\partial g_{\alpha\beta}}{\partial x^{\gamma}} 
\Bigr) = \Gamma^{\gamma}_{\beta\alpha}$ are the Cristoffel symbols of the second kind. It is interesting to 
note that the form of Eq.(\ref{Maxweq11}) does not depend explicitly upon the parameter $n$ which defines 
the dimension of Riemann space. By performing a few simple transformations we can reduce the formula, 
Eq.(\ref{Maxweq11a}), to the form which exactly coincides with Eq.(\ref{Maxweq1}). This has been noticed in 
many textbooks on three-dimensional electrodynamics (see, e.g., \cite{LLTF}).  

The second group of Maxwell equations for multi-dimensional spaces of non-zero curvature is written in the 
form (in Gauss units) 
\begin{eqnarray}
 \nabla_{\beta} F^{\alpha\beta} = \frac{1}{\sqrt{- g}} \frac{\Bigl(\partial \sqrt{- g} 
 F^{\alpha\beta}\Bigr)}{\partial x^{\beta}} = - \frac{n \pi^{\Bigl(\frac{n}{2}\Bigr)}}{c \Gamma\Bigl( 1 
 + \frac{n}{2} \Bigr)} \; j^{\alpha} = - \frac{f(n)}{c} j^{\alpha} \; \; , \; \label{Maxweq2a} 
\end{eqnarray}
since the tensor $F^{\alpha\beta}$ is antisymmetric. In this equation $g$ is the determinant of the fundamental 
tensor which is always negative in the metric gravity. By applying the operator $\nabla_{\alpha}$ to the last 
formula one finds
\begin{eqnarray}
 \nabla_{\alpha} \nabla_{\beta} F^{\alpha\beta} = - \frac{f(n)}{c} \nabla_{\alpha} j^{\alpha} \; \; 
 \Longrightarrow \; \; - \frac{f(n)}{c} \nabla_{\beta} j^{\beta} = \nabla_{\beta} \nabla_{\alpha} 
 F^{\beta\alpha} = - \nabla_{\beta} \nabla_{\alpha} F^{\alpha\beta} \; . \; \label{currentcon0} 
\end{eqnarray}
In other words, the expression in the left-hand side of this equation(s) can be re-written in the following 
form 
\begin{eqnarray}
 \frac12 \Bigl( \nabla_{\alpha} \nabla_{\beta} + \nabla_{\beta} \nabla_{\alpha} \Bigr) F^{\alpha\beta} \; . 
 \; \label{currentcon} 
\end{eqnarray}
which equals zero identically, since here the truly symmetric tensor operator (upon $\alpha \leftrightarrow 
\beta$ permutation) is applied to an antisymmetric tensor (upon the same permutations). Finally, one finds 
that $\nabla_{\alpha} j^{\alpha} = 0$, i.e., the conservation law for electric charge written in the $(n + 
1)-$dimensional Riemannian space.   
 
In many books and textbooks on electrodynamics derivation of Maxwell equations in the manifestly covariant 
form is traditionally considered as the final step. Similar approach, however, ignores an additional group 
of governing equations which are obeyed for the electromagnetic field in the presence of actual 
gravitational filed(s). These additional equations determine general properties, time-evolution and 
propagation of electromagnetic field(s) in the metric gravitational filed(s). Explicit derivation of these 
additional governing equations for the electromagnetic field tensor is straightforward. Indeed, if the 
electromagnetic field tensor $F_{\alpha\beta}$ is considered in the metric gravity, then the following 
equations must be obeyed 
\begin{eqnarray}
 \nabla_{\lambda} \nabla_{\sigma} F_{\alpha}^{\beta} - \nabla_{\sigma} \nabla_{\lambda} F_{\alpha}^{\beta} = 
 F^{\mu}_{\alpha} R^{\beta}_{\sigma\lambda\mu} - F^{\beta}_{\mu} R^{\mu}_{\sigma\lambda\alpha} 
 \; \; , \; \label{EquatA1} 
\end{eqnarray}
or in a slightly different form
\begin{eqnarray}
 \nabla_{\lambda} \nabla_{\sigma} F_{\alpha\beta} - \nabla_{\sigma} \nabla_{\lambda} F_{\alpha\beta} = 
 - F_{\mu\beta} R^{\mu}_{\sigma\lambda\alpha} - F_{\alpha\mu} R^{\mu}_{\sigma\lambda\beta} =
 F_{\alpha\mu} R^{\mu}_{\lambda\sigma\beta} + F_{\mu\beta} R^{\mu}_{\lambda\sigma\alpha} 
 \; \; , \; \label{EquatA2} 
\end{eqnarray}
where the notation $R^{\sigma}_{\alpha\beta \gamma} = g^{\sigma\mu} R_{\alpha\beta \gamma\mu}$ is the
Riemann-Cristoffel tensor of the fourth rank which is three times covariant and once contravariant (see, 
e.g., \cite{Kochin}, \cite{Dash}). In turn, the $R_{\alpha\beta \gamma\sigma}$ is the Riemann curvature 
tensor (or Riemann-Cristoffel tensor)
\begin{eqnarray}
 R_{\alpha\beta \gamma\sigma} = \frac12 \Bigl[ \frac{\partial^{2} g_{\alpha\sigma}}{\partial x^{\beta} 
 \partial  x^{\gamma}} + \frac{\partial^{2} g_{\beta\gamma}}{\partial x^{\alpha} \partial x^{\sigma}} 
  - \frac{\partial^{2} g_{\alpha\gamma}}{\partial x^{\beta} \partial x^{\sigma}}
  - \frac{\partial^{2} g_{\beta\sigma}}{\partial x^{\alpha} \partial x^{\gamma}} \Bigr] + \Gamma_{\rho, 
  \alpha\sigma} \Gamma^{\rho}_{\beta\gamma} - \Gamma_{\rho, \beta\sigma} \Gamma^{\rho}_{\alpha\gamma} \; 
  \; , \; \label{two} 
\end{eqnarray} 
where $\Gamma_{\gamma, \mu\nu} = \frac12 \Bigl( \frac{\partial g_{\gamma\alpha}}{\partial x^{\beta}} + 
\frac{\partial g_{\gamma\beta}}{\partial x^{\alpha}} - \frac{\partial g_{\alpha\beta}}{\partial x^{\gamma}} 
\Bigr)$ are the Cristoffel symbols of the first kind. The Riemann-Cristoffel tensor defined in Eq.(\ref{two}) 
is a covariant tensor of the fourth rank. Note that similar problems have been extensively studied since 
1920's in numerous papers and books on General Relativity (see, e.g., \cite{Tolm}, \cite{Carm} and references 
therein). As follows from these equations, Eqs.(\ref{EquatA1}) - (\ref{EquatA2}), propagation and other 
properties of the 'free' electromagnetic fields in multi-dimensional spaces of non-zero curvature (or in 
non-flat spaces) are always affected by the gravitational field(s). For relatively small gravitational field(s) 
Eqs.(\ref{EquatA1}) - (\ref{EquatA2}) can be considered as small perturbations to the Maxwell equations. However, 
in strong gravitational fields, where some of the $| \frac{\partial g_{\alpha\beta}}{\partial x^{\gamma}} |$ 
derivatives are very large, the laws of propagation and other properties of the electromagnetic field(s) can 
significantly be changed by the gravity. Briefly, we can say say that in similar non-flat spaces the actual 
properties of electromagnetic field(s) cannot be described by the Maxwell equations only. Furthermore, in more 
complex 'combined' theories of gravity and radiation, e.g., in the well known Born-Infeld theory (see, e.g. 
\cite{Raf}), the total fundamental tensor is represented as a function, e.g., as a sum, of the gravitational 
$g_{\alpha\beta}$ and electromagnetic $F_{\alpha\beta}$ tensors, the time-evolution and propagation of 
electromagnetic field(s) is described by the non-linear, well-coupled equations. 

\subsection{Multi-dimensional electromagnetic field in metric gravity}

Now, we are ready to vary the sum of action integrals for the gravitational $S_g$ and electromagnetic $S_f$ fields, 
i.e., to vary the $\delta (S_g + S_f)$ action. The both fields are considered as free, i.e., there are no masses, 
no free electric charges and no electric currents in the area of our interest. Our goal in this Section is to 
derive (variationally) the governing Einstein equations (in multi-dimensions) in the presence of electromagnetic 
field. To achieve this goal we have to vary the gravitational field only, i.e., the components of the metric tensor 
$g_{\alpha\beta}$ (or $g^{\alpha\beta}$). The variation of the gravitational action $S_g$ is written in the form 
(see, e.g., \cite{LLTF} and \cite{Carm})
\begin{eqnarray}
   \delta S_g = - \frac{c}{f(n) {\cal K}} \int \Bigl( R_{\alpha\beta} - \frac12 g_{\alpha\beta} R \Bigr)  
   \delta g^{\alpha\beta} \sqrt{- g} d\Omega \; , \; \label{EQ1}
\end{eqnarray} 
where $R_{\alpha\beta}$ is the Ricci tensor. In old books \cite{Kochin} they have used the the Einstein tensor
which is $G_{\alpha\beta} = - R_{\alpha\beta}$. The explicit form of the Ricci tensor is
\begin{eqnarray}
 R_{\alpha\beta} = \frac{\partial \Gamma^{\gamma}_{\alpha\beta}}{\partial x^{\gamma}} - \frac{\partial 
 \Gamma^{\gamma}_{\alpha\gamma}}{\partial x^{\beta}} + \Gamma^{\gamma}_{\alpha\beta} 
 \Gamma^{\lambda}_{\gamma\lambda} - \Gamma^{\lambda}_{\alpha\gamma} \Gamma^{\gamma}_{\beta\lambda} 
 \; , \; {\rm or} \; \; R_{\alpha\beta} = g^{\mu\nu} R_{\mu\alpha\beta\nu} = g^{\nu\mu} 
  R_{\nu\beta\alpha\mu} = R_{\beta\alpha} \; \label{equH1}  
\end{eqnarray}
and $R = g^{\alpha\beta} R_{\alpha\beta}$ is the scalar (or Gauss) curvature of space. Also in this equation 
the notation ${\cal K} = \frac{k}{c^2} = 7.4259155 \cdot 10^{-29}$ $cm \cdot sec^{-1}$ denotes the universal 
(or $n-$independent) gravitational constant. Similar variation of the electromagnetic action $S_f$ is 
\begin{eqnarray}
 \delta S_f = \frac{2}{c} \int T_{\alpha\beta} \delta g^{\alpha\beta} \sqrt{- g} d\Omega = \frac{2}{c f(n)} 
 \int \Bigl( F_{\alpha\gamma} F^{\gamma}_{\beta} + \frac14 g_{\alpha\beta} F_{\gamma\rho} F^{\gamma\rho} \Bigr) 
 \delta g^{\alpha\beta} \sqrt{- g} d\Omega \; . \; \label{EQ2} 
\end{eqnarray} 

Therefore, for the variation of the sum of these two actions we can write 
\begin{eqnarray}
 \delta ( S_g + S_f )= \frac{c}{f(n) {\cal K}} \int \Bigl( - R_{\alpha\beta} + \frac12 g_{\alpha\beta} R + 
 \frac{2 f(n) {\cal K}}{c^2} T_{\alpha\beta} \Bigr) \delta g^{\alpha\beta} \sqrt{- g} d\Omega \; . \; 
 \label{EQ3}
\end{eqnarray} 
Since variations of the gravitational field are arbitrary, then from this equation one finds 
\begin{eqnarray}
 R_{\alpha\beta} - \frac12 g_{\alpha\beta} R = \frac{2 {\cal  K}}{c^2} \Bigl( F_{\alpha\gamma} 
 F^{\gamma}_{\beta} + \frac14 g_{\alpha\beta} F_{\gamma\rho} F^{\gamma\rho} \Bigr) = 
 \frac{2 {\cal K}}{c^2} \tilde{T}_{\alpha\beta} \; \; , \; \label{EQ4} 
\end{eqnarray} 
where $\tilde{T}_{\alpha\beta} = F_{\alpha\gamma} F^{\gamma}_{\beta} + \frac14 g_{\alpha\beta} F_{\gamma\rho}
F^{\gamma\rho}$ is the reduced (or universal) energy-momentum tensor of the electromagnetic field which does 
not include the hyper-angular $f(n)$ factor. The last equation, Eq.(\ref{EQ4}), is the well known Einstein 
equation for the gravitational and electromagnetic field. This equation is a true tensor equation, since the 
both parts of this equation do not include the geometrical (or hyper-angular) factor $f(n)$. In other words, 
by looking at this equation one cannot say what is the actual dimension of our working space. For this reason 
Flanders \cite{Fland} and others have criticized the classical tensor analysis: ``In classical tensor analysis, 
one never knows what is the range of applicability simply because one is never told what the space is.''
However, for the purposes of this study this fact is an obvious advantage. Any of the true tensor equations, 
which appear in fundamental physics, cannot include factors which explicitly depend upon the dimension $n$ 
(or $n + 1$) of the working Riemann space. Moreover, this is a simple criterion which can be used to separate 
the true (also universal, or absolute) tensor equations from similar tensor-like equations which can be 
correct only for some selected Riemannian spaces. As follows from arguments presented above the both Einstein 
equations of the metric gravity for the free gravitational field, when $\tilde{T}_{\alpha\beta} = 0$ in 
Eq.(\ref{EQ4}), and Einstein equations of metric gravity in the presence of electromagnetic field, 
Eq.(\ref{EQ4}), are the true tensor equations. 

\subsection{Radiation from a rapidly moving electric charge}

As is well know (see, e.g., \cite{LLTF} and \cite{Jack}) any electric charge, which accelerates in the 
electromagnetic field, always emits EM-radiation. Nowadays, this statement is repeated so often that a 
large number of students and researchers sincerely believe that EM-radiation can only be emitted in the
presence of electromagnetic field. In general, this is not an absolute true and emission of EM-radiation 
is also possible in the presence of a strong (or rapidly varying) gravitational field. Below, we want to 
prove this statement and, for simplicity, here we restrict our analysis to the three-dimensional space 
only. However, all our formulas will be written in the explicitly covariant form. This means that all 
these formulas can be generalized to describe the actual situation in multi-dimensional spaces too. In 
general relativity the formula for the radiated four-momentum $d P^{\kappa}$ is written in the form (see, 
e.g., \cite{LLTF})   
\begin{eqnarray}
 d P^{\kappa} = - \frac{2 e^{2}}{3 c} g_{\alpha\mu} \Bigl(\frac{d^{2} x^{\alpha}}{d s^{2}}\Bigr) 
 \Bigl(\frac{d^{2} x^{\mu}}{d s^{2}}\Bigr) d x^{\kappa} = - \frac{2 e^{2}}{3 c} g_{\alpha\mu} 
 \Bigl(\frac{d u^{\alpha}}{d s}\Bigr) \Bigl(\frac{d u^{\mu}}{d s}\Bigr) u^{\kappa} ds \; \; , 
 \; \; \label{ForRad1} 
\end{eqnarray} 
where $u^{\beta} = \frac{d x^{\beta}}{d s}$ is the corresponding `velocity'. Now, by taking the expression 
for the acceleration from, Eq.(\ref{eqamot}), one finds  
\begin{eqnarray}
 &&d P^{\kappa} = - \frac{2 e^{2}}{3 c} g_{\alpha\mu} \Bigl( \Gamma^{\alpha}_{\beta\gamma} u^{\beta} 
 u^{\gamma} - \frac{e}{m c^{2}} F^{\alpha}_{\beta} u^{\beta} \Bigr) \Bigl( \Gamma^{\mu}_{\lambda\sigma} 
 u^{\lambda} u^{\sigma} - \frac{e}{m c^{2}} F^{\mu}_{\sigma} u^{\sigma} \Bigr) u^{\kappa} ds = 
 - \frac{2 e^{2}}{3 c} \times \nonumber \\
 &&\Bigl( g_{\alpha\mu} \Gamma^{\alpha}_{\beta\gamma} \Gamma^{\mu}_{\lambda\sigma} u^{\beta} u^{\gamma} 
 u^{\lambda} u^{\sigma} u^{\kappa} - \frac{2 e}{m c^{2}} g_{\alpha\mu} \Gamma^{\alpha}_{\beta\gamma} 
 F^{\mu}_{\sigma} u^{\beta} u^{\gamma} u^{\sigma} u^{\kappa} + \frac{e^{2}}{m^{2} c^{4}} g_{\alpha\mu} 
 F^{\alpha}_{\beta} F^{\mu}_{\sigma} u^{\beta} u^{\sigma} u^{\kappa} \Bigr) \; \; \label{ForRad2} \\
 &&= T^{\kappa}_{2} + T^{\kappa}_{2} + T^{\kappa}_{3} =- \frac{2 e^{2}}{3 c} 
 \Gamma^{\alpha}_{\beta\gamma} \Gamma_{\alpha,\lambda\sigma} u^{\beta} u^{\gamma} u^{\lambda} 
 u^{\sigma} u^{\kappa} + \frac{4 e^{3}}{3 m c^{3}} \Gamma^{\alpha}_{\beta\gamma} F_{\alpha\sigma} 
 u^{\beta} u^{\gamma} u^{\sigma} u^{\kappa} \nonumber \\
 &&- \frac{2 e^{4}}{3 m^{2} c^{5}} F^{\alpha}_{\beta} F_{\alpha\sigma} u^{\beta} u^{\sigma} u^{\kappa} 
 \; \; , \; \; \nonumber 
\end{eqnarray} 
where the last term (vector) $T^{\kappa}_{3} = - \frac{2 e^{4}}{3 m^{2} c^{5}} F^{\alpha}_{\beta} 
F_{\alpha\sigma} u^{\beta} u^{\sigma} u^{\kappa}$. This term describes the emission of EM-radiation 
by a single electrical charge which is rapidly moving in some electromagnetic field. It was extensively
discussed in numerous books on classical electrodynamics (see, e.g., \cite{LLTF} and \cite{Jack}) and 
below we do not want to repeat these discussions. The first term in Eq.(\ref{ForRad2}) $T^{\kappa}_{1} 
= - \frac{2 e^{2}}{3 c} \Gamma^{\alpha}_{\beta\gamma} \Gamma_{\alpha,\lambda\sigma} u^{\beta} u^{\gamma} 
u^{\lambda} u^{\sigma} u^{\kappa}$ is also a vector. This vector represents the emission of EM-radiation 
by a point electric charge which rapidly moves in the gravitational field. The second term (vector) in 
Eq.(\ref{ForRad2}) describes the interference between gravitational and electromagnetic emissions of the 
EM-field. The explicit formula for this term is $T^{\kappa}_{2} = \frac{4 e^{3}}{3 m c^{3}} 
\Gamma^{\alpha}_{\beta\gamma} F_{\alpha\sigma} u^{\beta} u^{\gamma} u^{\sigma} u^{\kappa}$. 

There are a number of interesting observations which directly follow from the three formulas for the 
$T^{\kappa}_{1}, T^{\kappa}_{2}$ and $T^{\kappa}_{3}$ terms in Eq.(\ref{ForRad2}). First, let us note 
that the $T^{\kappa}_{1}$ term does not contain any particle mass. This means that one fast electron 
and/or one fast proton, which move with the equal velocities in a pure gravitational field, will always 
emit equal amount of radiation. This the main distinguishing feature of the gravitation emission of 
EM-radiation. Second, this term is a fifth-order power function of the velocities. Therefore, it 
is clear that overall contribution of this term will rapidly increase for relativistic particles which 
move with the velocities close to the speed of light in vacuum $c$. It is also clear that usually in 
Eq.(\ref{ForRad2}) the third term $T^{\kappa}_{3}$ is substantially larger than two other terms. 
In other words, the gravitational emission of EM-radiation is hard to observe at `normal' gravitational 
conditions. However, in strong gravitational fields, where the absolute values of Cristoffel symbols 
are very large (or the $| \frac{\partial g_{\alpha\beta}}{\partial x^{\gamma}} |$ derivatives are very 
large) the situation can be different. The second condition is simple: the rapidly moving particle must 
be truly relativistic, i.e., it must move with the velocity which is close to the speed of light $v \ge 
0.9 \; c$ in respect to the system where the rapidly changing gravitational field was originated. If 
these two conditions are obeyed, then one can see a relatively intense gravitational EM-radiation which 
is emitted by a single relativistic particle which has non-zero electric charge $e$. 

\section{Conclusions}

We have generalized the three-dimensional Maxwell theory of radiation to multi-dimensional flat and 
curved spaces. Some equations derived in three-dimensional Maxwell electrodynamics do not change 
their form in multi-dimensional space. In other equations we have to make a number of changes. In 
fact, all properties of the electromagnetic field are described by the $(n + 1)$-dimensional vector 
potential $\bar{A} = (\phi, {\bf A})$, while interaction between any particle and electromagnetic 
field are described by one experimental parameter, which is the electric charge $e$ of this particle. 
The governing Maxwell equations for the multi-dimensional electromagnetic field have been derived and 
written in the covariant (or tensor) form. These equations include the geometrical (or hyper-angular) 
factor $f(n) = \frac{n \pi^{\frac{n}{2}}}{\Gamma\Bigl( 1 + \frac{n}{2} \Bigr)}$, which explicitly 
depend upon the dimension of space $n$. 

The Hamiltonian formulation of the Maxwell radiation field in multi-dimensional spaces is developed and 
investigated. We have found that the total number of first-class constraints in this Hamiltonian 
formulation equals two (one primary and one secondary constraints). This number exactly coincides with 
the number of first-class constraints in analogous Hamiltonian formulation developed earlier by Dirac 
\cite{Dir64} for the pure radiation field in three-dimensional space. In other words, the total number 
of first-class constraints in any Hamiltonian formulation developed for the free radiation field does 
not depend upon the dimension of space $n$. To understand how lucky we are with the Hamiltonian 
formulations of electrodynamics, let us note that in the $(n + 1)-$dimensional metric gravity we always 
have $(n + 1)$ primary and $(n + 1)$ secondary first-class constraints. In addition to this, in many 
sets of canonical variables the explicit form of all arising secondary constraints are very cumbersome 
(see, e.g., \cite{Dir58} - \cite{Fro2020}) and this substantially complicates all operations with these 
values. By using these primary and secondary first class constraints we have investigated the gauge 
conditions in multi-dimensional electrodynamics. 

Also, in the last Section the Maxwell equations in multi-dimensional non-flat spaces are written in the 
manifestly covariant form. It is shown that any gravitation field changes the actual properties, 
time-evolution and space-time propagation of electromagnetic field(s). For gravitation fields with large 
and very large connectivity coefficients $\Gamma^{\alpha}_{\beta\gamma}$ the 'pure' radiation field 
cannot be described by the Maxwell equations only. Additional equations for the antisymmetric tensor of
the electromagnetic field $F_{\alpha\beta}$ (and $F_{\alpha}^{\beta}$) has been derived in this study
(see, Eqs.(\ref{EquatA1}) and (\ref{EquatA2})). Analogous equation for the reduced energy-momentum tensor 
of electromagnetic field is now written in the true tensor form (see, Eq.(\ref{EQ4})), which does not 
contain any $n-$dependent factor. 

In conclusion, we wish to note that investigation of multi-dimensional Maxwell equations is not a pure 
academic problem. In fact, there are a number of advantages which one can gain by performing such an 
investigation. First, it does helps to clarify additional and interesting features of Maxwell's 
equations in the usual three-dimensional space (or in four-dimensional space-time). By working only 
with the three-dimensional Maxwell equations in our everyday life we simply do not pay attention on 
some fundamental and amazing facts. Second, if we have a complete and correct formulation for Maxwell's 
electrodynamics in multi-dimensional spaces, then it possible to develop the so-called unified theories 
of various fields, which include an electromagnetic field. In particular, the correct unified theory of 
the gravitational and electromagnetic fields in multi-dimensional spaces is of great interest in modern
theoretical physics. Third, recently in experiments in high-energy physics it has been noted that at very 
high collision energies many results can be represented to very good numerical accuracy and with higher 
symmetry, if we introduce multi-dimensional spaces at the intermediate stages of calculations. This fact 
is not completely unexpected, but we need to understand the internal nature of such a phenomenon. If 
multi-dimensional spaces do play a significant role during such processes, then it can change a lot in 
modern physics and natural philosophy. Note that some of the problems mentioned in this study have been 
considered earlier (see, e.g., \cite{2015} - \cite{2013}).  \\

\appendix
\section{Scalar electrodynamics}
\label{A} 

In this study our analysis of electrodynamics in multi-dimensional spaces was restricted to the spaces
which have geometrical dimension $n \ge 3$. For the sake of completeness, now we want to consider the 
one- and two-dimensional spaces. To investigate these small-dimensional cases we shall apply one 
effective method which is based on the so-called scalar electrodynamics. This `pre-Maxwell' method was 
described and briefly discussed in \cite{Fro2015}. Scalar electrodynamics can be introduced in 
three-dimensional space where one can compare the arising equations with the usual Maxwell equations. 
The foundation of scalar electrodynamics is the well known theorem from vector calculus (see, e.g., 
\cite{Kochin}) which stays that an arbitrary vector ${\bf B}$ in three-dimensional space can be 
represented in the following $two-gradient$ form 
\begin{eqnarray} 
  {\bf B} = \Psi_1 \nabla \Psi_2 + \nabla \Psi_3  \; \; , \; \; \label{SEeq1} 
\end{eqnarray}
where $\Psi_1, \Psi_2$ and $\Psi_3$ three arbitrary analytical functions of three spatial and one 
temporal coordinates. In general, each of these functions can be real, or complex. This formula 
can directly be applied to the vector potential of the electromagnetic field ${\bf A}$. The 
four-dimensional vector potential $(\varphi, {\bf A})$ and intensities of electric ${\bf E}$ and 
magnetic ${\bf H}$ field are also represented in terms of the four $\Psi_1, \Psi_2, \Psi_3$ and 
$\varphi$ scalar functions. For two- and one-dimensional (geometrical) spaces the total number of 
such scalar functions equals three and two, respectively. 

To derive the explicit expressions and obtain the governing equations of electrodynamics one needs 
to use the two following formulas which play a central role in scalar electrodynamics
\begin{eqnarray}
 curl {\bf A} = \nabla \Psi_1 \times \nabla \Psi_2 \; \; \; \; {\rm and} \; \; \; \;  div {\bf A} = 
 \Psi_1 \Delta \Psi_2 + \nabla \Psi_1 \cdot \nabla \Psi_2 + \Delta \Psi_3 \; \; \label{SEeq3} 
\end{eqnarray} 
As follows from Eq.(\ref{SEeq3}) in scalar electrodynamics there are a number of advantages to choose 
some of the $\Psi_1, \Psi_2$ and $\Psi_3$ functions (where it is possible) as harmonic functions for 
which $\Delta \Psi_k = 0$, where $k$ = 1, 2, 3. Such a choice of functions reduces the total number of 
terms in Maxwell equations and gauge conditions. In turn, this simplify analysis and solutions of many
problems in scalar electrodynamics. In fact, in three-dimensional spaces the scalar electrodynamics 
cannot compete with the traditional vector approach. The main reason is obvious, since the regular 
Maxwell equations are linear for all components of the electromagnetic field. However, some selected 
three-dimensional problems can be solved (completely and accurately), if we apply the method of scalar 
electrodynamics. 

For two-dimensional spaces equation, Eq.(\ref{SEeq1}), take the form: ${\bf A} = \Psi_1 \nabla \Psi_2$, 
since in this case we can assume that $\Psi_3 = 0$. The equality ${\bf A} \cdot curl {\bf A} = 0$ is a 
necessary and sufficient condition to represent the vector ${\bf A}$ in such a form \cite{Kochin} (it 
does obey in this case). This leads to the following equations: 
\begin{eqnarray}
 {\bf H} =  curl {\bf A} = \nabla \Psi_1 \times \nabla \Psi_2 \; \; \; \; {\rm and} \; \; \; \;  
 div {\bf A} = \Psi_1 \Delta \Psi_2 + \nabla \Psi_1 \cdot \nabla \Psi_2 \; \; . \; \label{SEeq4} 
\end{eqnarray} 
We also need the explicit expression for the $curl {\bf H}$ 
\begin{eqnarray}
 curl {\bf H} = \nabla \Psi_1 \Delta \Psi_2 - \nabla \Psi_2 \Delta \Psi_1 + (\nabla \Psi_1 \cdot 
 \nabla) \Psi_2 - (\nabla \Psi_2 \cdot \nabla) \Psi_1 = \nabla \Psi_1 \Delta \Psi_2 - \nabla \Psi_2 
 \Delta \Psi_1 \; \; \nonumber 
\end{eqnarray} 
Also note that if $\Psi_2$ is chosen as a harmonic function, i.e., $\Delta \Psi_2 = 0$, and $\nabla 
\Psi_1 \perp \nabla \Psi_2$, then the gauge condition is obeyed automatically and solutions of a 
large number of problems known in two-dimensional electrodynamics simplifies significantly. In 
general, it can be shown that the both two-dimensional electrodynamics and two-dimensional 
electrostatics include a number of operations with the harmonic functions (see, e.g., \cite{TikhS} - 
\cite{LLES}). In turn, this leads to numerous successful applications of conformal mapping methods 
to describe the two-dimensional electromagnetic waves and determine solutions of numerous problems in 
two-dimensional electrostatics. 

In one dimensional case from equation, Eq.(\ref{SEeq1}), one finds ${\bf A} = \nabla \Psi_2 = \nabla 
\Psi$. Therefore, the $curl$ of the vector potential equals zero identically. This means that there 
is no classical magnetic field in one-dimensional space. Moreover, any time-variations of the electric 
field cannot generate any magnetic field, i.e., we have no Faraday's law in one-dimensional 
(geometrical) space. In other words, the one-dimensional electrodynamics does not exist. On the other 
hand, many one-dimensional electrostatic problems which include the potential and intensity of the 
electric field only, can still be formulated and solved correctly. \\


\begin{thebibliography}{99}

\bibitem{Fro2021} A.M. Frolov, Physics of Atomic Nuclei (Yad. Fiz.), {\bf 84}(5), 750 (2021). 

\bibitem{Maxw} J.C. Maxwell, Phil. Trans. Royal Society of London, {\bf 155}, 459 (1865).

\bibitem{Maxw1} J.C. Maxwell, {\it A Treatise on Electricity and Magnetism}, in 2 vols. (Oxford University Press, 
Oxford, 1873).

\bibitem{Fro2015} A.M. Frolov, {\it On the 150th anniversary of Maxwell equations}, Journal of Multidisciplinary 
Engineering Science and Technology (JMEST), {\bf 2}, 361 (2015). 

\bibitem{LLTF} L.D. Landau and E.M. Lifshitz, {\it The Classical Theory of Fields}, 4th ed. (Pergamon Press Ltd., 
London, 1975). 

\bibitem{Kochin} N. E. Kochin, \textit{Vector Calculus and the Principles of Tensor Calculus}, 9th ed. (USSR Acad. 
of Sciences Publishing, Moscow, 1965).

\bibitem{Dash} P.K. Dashevskii, \textit{Riemannian Geometry and Tensor Analysis}, 3rd. ed. (Nauka, 
Moscow, 1967).

\bibitem{Sok1} A.A. Sokolov, \textit{Introduction in Quantum Electrodynamics}, (Fizmatgiz, Moscow, 1958) [in 
Russian]. 

\bibitem{Sok2} A.A. Sokolov and I.M. Ternov, \textit{Relativistic electron}, 2nd ed. (Nauka, Moscow),
(1983), Chpts. 1 and 2 [in Russian].  

\bibitem{GR} I.S. Gradstein and I.M. Ryzhik, \textit{Tables of Integrals, Series and Products}, 6th revised ed. 
(Academic Press, New York, 2000).

\bibitem{Fland} H. Flanders, {\it Differential Forms with Applications to the Physical Sciences}, (Dover 
Publications, Inc., Mineola, New York, 1989). 

\bibitem{Dir48} P. A. M. Dirac, Physical Review {\bf 74}, 817 (1948). 

\bibitem{Amaldi} E. Amaldi, \textit{On the Dirac Magnetic Poles}, in: {\it Old and New Problems in Elementary 
Particles}, Ed. G. Puppi, (Academic Press, New York, 1968).

\bibitem{Dir64} P. A. M. Dirac, \textit{Lectures on Quantum Mechanics}, (Befler Graduate School of 
Sciences, Yeshiva University, New York, 1964).

\bibitem{Dir50} P. A. M. Dirac, Canadian Journal of Mathematics {\bf 2}, 129 (1950).

\bibitem{Cast} L. Castellani, Annals of Physics {\bf 143}, 357 (1982).

\bibitem{BHag} E. Bessel-Hagen, Mathematische Annalen {\bf 84}, 258 (1921). 

\bibitem{Heitl} W. Heitler, \textit{The Quantum Theory of Radiation}, 3rd ed. (Oxford University Press, 
London, UK, 1954). 

\bibitem{Jack} J.D. Jackson, \textit{Classical Electrodynamics}, 2nd. ed. (J. Wiley \& Sons Inc., New York, 1975), 
Sect. 6.5

\bibitem{Jack1} J.D. Jackson, American Journal of Physics {\bf 70}, 917 (2002).  

\bibitem{GF} I. M. Gelfand and S. V. Fomin, \textit{Calculus of Variations}, (Dover Publ., Inc., Mineola, New York, 1990), 
Chpt. 7. 

\bibitem{Wol} W. Engelhardt, Annals de la Fondation Louis de Broglie, {\bf 30}, 157 (2005).  

\bibitem{Tolm} R.C. Tolmen, \textit{Relativity, Thermodynamics and Cosmology}, 3rd ed. (Oxford at the Clarendon Press, 
Oxford, UK, 1969), Chpt. VIII. 
 
\bibitem{Carm} M. Carmeli, \textit{Classical Fields: General Relativity and Gauge Theory} (World Scientific Publ. Co., 
Singapore, 2002). 
 
\bibitem{Raf} J. Rafelski, W. Greiner and L.P. Fulcher, Nuovo Cimento {\bf 13 B}, 135 (1973).

\bibitem{Dir58} P. A. M. Dirac, Proceedings of the Royal Society {\bf 246}, 333 (1958).

\bibitem{K&K} N. Kiriushcheva, S. V. Kuzmin, C. Racnkor, and S. R. Valluri, Physics Letters A {\bf 372}, 5101 (2008).

\bibitem{FK&K} A. M. Frolov, N. Kiriushcheva, and S. V. Kuzmin, Gravitation and Cosmology {\bf 17}, 314 (2011). 

\bibitem{Fro2020} A. M. Frolov, Canadian Journal of Physics {\bf 98}, 405 (2020).

\bibitem{2015} P. Lemos, G.M. Quinta and O.B. Zaslavskii, Phys. Rev. D {\bf 91}, 104027 (2015). 

\bibitem{2018} P.V. Kratovich and Ju.V. Tchemarina, arXiv: 1805.02698 (2018). 

\bibitem{2019} Z. Yousaf, K. Bamba and U. Ghafoor, Physical Review D {\bf 100}, 024062 (2019). 

\bibitem{2013} D. Pugliese and J.A. Valiente Kroon, Relativity and Gravitation {\bf 45}, 1247 (2013). 


\bibitem{TikhS} A.N. Tikhonov and A.A. Samarskii, {\it Equations of Mathematical Physics}, (Dover Publ., Inc., 1990),
Chpt. IV. 

\bibitem{Smythe} W.R. Smythe, {\it Static and Dynamic Electricity}, (McGraw-HIll, Inc., New York, 1950).

\bibitem{LLES} L.D. Landau and E.M. Lifshitz, {\it Electrodynamics of Continuous Media}, 2nd revised ed. 
(Pergamon Press Ltd., London, 1984). 

\end{thebibliography}
\end{document}